%% file: main.tex
\documentclass[12pt,a4paper]{article}

\usepackage{ifthen} 
\usepackage{pgffor} 
\usepackage{enumerate}
\usepackage{fancyvrb}
\usepackage{hyperref}
\usepackage{lineno}
\newboolean{pdflatex}
\setboolean{pdflatex}{true} 

\newboolean{articletitles}
\setboolean{articletitles}{true} 

\newboolean{uprightparticles}
\setboolean{uprightparticles}{false} 

\newboolean{inbibliography}
\setboolean{inbibliography}{false} 

\input{preamble}
\usepackage{longtable} 

\def\figwidth{0.45}
\def\Zmm     {{\ensuremath{\Z \to \mumu}}\xspace}

\def\Wpmn     {{\ensuremath{\Wp \to \mup\neum}}\xspace}
\def\Wmmn     {{\ensuremath{\Wm \to \mun\neumb}}\xspace}
\def\mum  {\ensuremath{{\,\upmu\mathrm{m}}}\xspace}

\def\stw  {{\ensuremath{\sin^2\theta^{\text{eff}}_{\text{lept}}}}\xspace}
\begin{document}

\renewcommand{\thefootnote}{\fnsymbol{footnote}}
\setcounter{footnote}{1}

\input{title-LHCb-ANA}

\renewcommand{\thefootnote}{\arabic{footnote}}
\setcounter{footnote}{0}

\tableofcontents


\pagestyle{plain} 
\setcounter{page}{1}
\pagenumbering{arabic}


\input{Section__Introduction}
\input{Section__Sample}
\input{Section__Method}
\input{Section__Validation}
\input{Section__Application}
\input{Section__MockUpMeasurements}
\input{Section__Conclusion}\clearpage
\input{Section__Acknowledgements}

\addcontentsline{toc}{section}{References}
\setboolean{inbibliography}{true}
\bibliographystyle{LHCb}
\bibliography{mybib}

\end{document}

%% file: preamble.tex
\usepackage{microtype}
\usepackage{lineno}  
\usepackage{xspace} 
\usepackage{caption} 

\usepackage{graphicx}  
\DeclareGraphicsExtensions{{.pdf,.png}}
\usepackage{color}
\usepackage{colortbl}
\graphicspath{{../plots/}} 

\usepackage{amsmath} 
\usepackage{amssymb}
\usepackage{amsfonts}
\usepackage{upgreek} 

\newcommand*\patchAmsMathEnvironmentForLineno[1]{%
\expandafter\let\csname old#1\expandafter\endcsname\csname #1\endcsname
\expandafter\let\csname oldend#1\expandafter\endcsname\csname
end#1\endcsname
 \renewenvironment{#1}%
   {\linenomath\csname old#1\endcsname}%
   {\csname oldend#1\endcsname\endlinenomath}%
}
\newcommand*\patchBothAmsMathEnvironmentsForLineno[1]{%
  \patchAmsMathEnvironmentForLineno{#1}%
  \patchAmsMathEnvironmentForLineno{#1*}%
}
\AtBeginDocument{%
\patchBothAmsMathEnvironmentsForLineno{equation}%
\patchBothAmsMathEnvironmentsForLineno{align}%
\patchBothAmsMathEnvironmentsForLineno{flalign}%
\patchBothAmsMathEnvironmentsForLineno{alignat}%
\patchBothAmsMathEnvironmentsForLineno{gather}%
\patchBothAmsMathEnvironmentsForLineno{multline}%
\patchBothAmsMathEnvironmentsForLineno{eqnarray}%
}

\usepackage{hyperref}    
\usepackage[all]{hypcap} 

\input{lhcb-symbols-def} 

\usepackage{cite} 
\usepackage{mciteplus}

%% file: lhcb-symbols-def.tex
\usepackage{xspace} 
\usepackage{upgreek}

\def\MagUp {\mbox{\em Mag\kern -0.05em Up}\xspace}

\ifthenelse{\boolean{uprightparticles}}%
{

 \def\Pmu         {\ensuremath{\upmu}\xspace}                 
 \def\Pnu         {\ensuremath{\upnu}\xspace}

 \def\PDelta      {\ensuremath{\Delta}\xspace}                 
 \def\PXi      {\ensuremath{\Xi}\xspace}                 
 \def\PLambda      {\ensuremath{\Lambda}\xspace}                 
 \def\PSigma      {\ensuremath{\Sigma}\xspace}                 
 \def\POmega      {\ensuremath{\Omega}\xspace}                 
 \def\PUpsilon      {\ensuremath{\Upsilon}\xspace}                 
 
 \def\PB      {\ensuremath{\mathrm{B}}\xspace}                 
                  
 \def\PD      {\ensuremath{\mathrm{D}}\xspace}

 \def\PK      {\ensuremath{\mathrm{K}}\xspace}

 \def\PW      {\ensuremath{\mathrm{W}}\xspace}

 \def\PZ      {\ensuremath{\mathrm{Z}}\xspace}

 \def\Pi      {\ensuremath{\mathrm{i}}\xspace}

}
{

 \def\Pmu         {\ensuremath{\mu}\xspace}                 
 \def\Pnu         {\ensuremath{\nu}\xspace}

 \mathchardef\PDelta="7101
 \mathchardef\PXi="7104
 \mathchardef\PLambda="7103
 \mathchardef\PSigma="7106
 \mathchardef\POmega="710A
 \mathchardef\PUpsilon="7107
                  
 \def\PB      {\ensuremath{B}\xspace}                 
                  
 \def\PD      {\ensuremath{D}\xspace}

 \def\PK      {\ensuremath{K}\xspace}

 \def\PW      {\ensuremath{W}\xspace}

 \def\PZ      {\ensuremath{Z}\xspace}

 \def\Pi      {\ensuremath{i}\xspace}

}

\makeatletter
\ifcase \@ptsize \relax
  \newcommand{\miniscule}{\@setfontsize\miniscule{4}{5}}
\or
  \newcommand{\miniscule}{\@setfontsize\miniscule{5}{6}}
\or
  \newcommand{\miniscule}{\@setfontsize\miniscule{5}{6}}
\fi
\makeatother

\DeclareRobustCommand{\optbar}[1]{\shortstack{{\miniscule (\rule[.5ex]{1.25em}{.18mm})}
  \\ [-.7ex] $#1$}}

\def\mup        {{\ensuremath{\Pmu^+}}\xspace}
\def\mun        {{\ensuremath{\Pmu^-}}\xspace} 
\def\mumu       {{\ensuremath{\Pmu^+\Pmu^-}}\xspace}

\def\neu        {{\ensuremath{\Pnu}}\xspace}
\def\neub       {{\ensuremath{\overline{\Pnu}}}\xspace}
\def\neum       {{\ensuremath{\neu_\mu}}\xspace}
\def\neumb      {{\ensuremath{\neub_\mu}}\xspace}

\def\Wp     {{\ensuremath{\PW^+}}\xspace}
\def\Wm     {{\ensuremath{\PW^-}}\xspace}

\def\Z      {{\ensuremath{\PZ}}\xspace}

  \def\Kbar    {{\kern 0.2em\overline{\kern -0.2em \PK}{}}\xspace}

\def\KorKbar    {\kern 0.18em\optbar{\kern -0.18em K}{}\xspace}

  \def\Dbar    {{\kern 0.2em\overline{\kern -0.2em \PD}{}}\xspace}

\def\DorDbar    {\kern 0.18em\optbar{\kern -0.18em D}{}\xspace}

\def\Bbar    {{\ensuremath{\kern 0.18em\overline{\kern -0.18em \PB}{}}}\xspace}

\def\BorBbar    {\kern 0.18em\optbar{\kern -0.18em B}{}\xspace}

  \def\Y#1S{\ensuremath{\PUpsilon{(#1S)}}\xspace}

\def\Lbar        {{\ensuremath{\kern 0.1em\overline{\kern -0.1em\PLambda}}}\xspace}
\def\LorLbar    {\kern 0.18em\optbar{\kern -0.18em \PLambda}{}\xspace}

\def\to                 {\ensuremath{\rightarrow}\xspace}

\def\stw     {{\ensuremath{\sin\theta_{\mathrm{W}}}}\xspace}

\def\AT#1     {\ensuremath{A_{\mathrm{T}}^{#1}}\xspace}           

\def\C#1      {\ensuremath{\mathcal{C}_{#1}}\xspace}                       
\def\Cp#1     {\ensuremath{\mathcal{C}_{#1}^{'}}\xspace}                    
\def\Ceff#1   {\ensuremath{\mathcal{C}_{#1}^{\mathrm{(eff)}}}\xspace}        
\def\Cpeff#1  {\ensuremath{\mathcal{C}_{#1}^{'\mathrm{(eff)}}}\xspace}       
\def\Ope#1    {\ensuremath{\mathcal{O}_{#1}}\xspace}                       
\def\Opep#1   {\ensuremath{\mathcal{O}_{#1}^{'}}\xspace}                    

\newcommand{\tev}{\ifthenelse{\boolean{inbibliography}}{\ensuremath{~T\kern -0.05em eV}\xspace}{\ensuremath{\mathrm{\,Te\kern -0.1em V}}}\xspace}
\newcommand{\gev}{\ensuremath{\mathrm{\,Ge\kern -0.1em V}}\xspace}
\newcommand{\mev}{\ensuremath{\mathrm{\,Me\kern -0.1em V}}\xspace}
\newcommand{\kev}{\ensuremath{\mathrm{\,ke\kern -0.1em V}}\xspace}
\newcommand{\ev}{\ensuremath{\mathrm{\,e\kern -0.1em V}}\xspace}
\newcommand{\gevc}{\ensuremath{{\mathrm{\,Ge\kern -0.1em V\!/}c}}\xspace}
\newcommand{\mevc}{\ensuremath{{\mathrm{\,Me\kern -0.1em V\!/}c}}\xspace}
\newcommand{\gevcc}{\ensuremath{{\mathrm{\,Ge\kern -0.1em V\!/}c^2}}\xspace}
\newcommand{\gevgevcccc}{\ensuremath{{\mathrm{\,Ge\kern -0.1em V^2\!/}c^4}}\xspace}
\newcommand{\mevcc}{\ensuremath{{\mathrm{\,Me\kern -0.1em V\!/}c^2}}\xspace}

\def\mum  {\ensuremath{{\,\upmu\mathrm{m}}}\xspace}

\def\gsim{{~\raise.15em\hbox{$>$}\kern-.85em
          \lower.35em\hbox{$\sim$}~}\xspace}
\def\lsim{{~\raise.15em\hbox{$<$}\kern-.85em
          \lower.35em\hbox{$\sim$}~}\xspace}

\def\tell1  {TELL1\xspace}
\def\ukl1   {UKL1\xspace}



%% file: title-LHCb-ANA.tex

\begin{titlepage}

\vspace*{-1.5cm}

\noindent
\begin{tabular*}{\linewidth}{lc@{\extracolsep{\fill}}r@{\extracolsep{0pt}}}
 \\
 & & \\ 
\hline
\end{tabular*}

\vspace*{4.0cm}

{\normalfont\bfseries\boldmath\Large
\begin{center}
A simple method to determine charge-dependent curvature biases in track reconstruction in hadron collider experiments
\end{center}
}

\vspace*{2.0cm}

\begin{center}
William Barter$^1$, Martina Pili$^2$, Mika Vesterinen$^3$
\bigskip\\
{\normalfont\itshape\footnotesize
$ ^1$Imperial College London, London, United Kingdom\\  
$ ^2$University of Oxford, Oxford, United Kingdom\\
$ ^3$University of Warwick, Coventry, United Kingdom\\
}
\end{center}

\vspace{\fill}

\begin{abstract}
  \noindent
  A new data-driven method, using $Z\to\mu^+\mu^-$ decays, is proposed to correct for charge-dependent curvature biases in spectrometer
  experiments at hadron colliders. 
  The method is studied assuming a detector with a
  ``forward-spectrometer" geometry similar to that of the LHCb experiment,
  and is shown to reliably control several simplified detector mis-alignment configurations.
  The applicability of the method for use in  measurements of precision electroweak observables is evaluated. 
 \end{abstract}

\vspace*{2.0cm}
\vspace{\fill}

\end{titlepage}

\pagestyle{empty}  


\newpage
\setcounter{page}{2}
\mbox{~}


%% file: Section__Introduction.tex
\section{Introduction}
\label{sec:intro}
The measurement of charged particle momenta in hadron collider experiments is susceptible to mis-alignments, inaccuracies in the knowledge of the magnetic field, or other biases in the reconstruction algorithms.
Precision measurements of electroweak parameters such as the $W$ boson mass ($m_W$) and the weak
mixing angle (\stw), using muonic decays of weak bosons, are particularly sensitive to 
the accuracy with which these details are modelled in simulations of the signal processes.
In this paper we focus on curvature biases of the type
 \begin{equation}
    \frac{q}{p} \rightarrow \frac{q}{p} + \delta,\label{eq:curvature}
\end{equation}
 for particles of charge $q$ and momentum $p$ driven, for example, by weak modes in the detector alignment procedure.
It is desirable to identify and eliminate these ($\delta$) biases from both the real and simulated data,
so that the simulation can be tuned and validated with better reliability.
 
In the first measurement of $m_W$ with the ATLAS experiment~\cite{ATLASmW}, track curvature biases were determined using the ratio of the energy and momentum of electrons from $W\to e\nu$ decays.
This approach is only applicable for detectors with comparable kinematic resolution for electrons and muons. A second approach, based on the determination of \emph{sagitta} biases in \Zmm decays was also studied.
The dependence of the invariant mass of \Zmm decays on the muon kinematics is sensitive to curvature biases but, particularly in the case of global mis-alignments, the biases on the $\mu^+$ and $\mu^-$ momenta can be strongly anti-correlated.
Ref.~\cite{Bodek} details a method in which the curvature biases are determined by assuming that the 
 mean $1/p_T$ of the $\mu^+$ and $\mu^-$ should follow the expectation of simulation, where $p_T$ is the momentum of the muon transverse to the beam axis. 
 The reliance on simulation arises because the $p_T$ distributions are sensitive to both mis-alignment \emph{and} physics effects (including intrinsic differences between the $\mu^+$ and $\mu^-$ kinematic distributions in $Z$ decays).
 That method has nevertheless been successfully applied in, for example, measurements of \stw by CDF~\cite{Aaltonen:2014loa} and CMS~\cite{CMSs2tw}.

 This paper presents an alternative data-driven approach to determine charge-dependent curvature bias corrections using \Zmm decays.
 The method is validated with the example of the LHCb experiment~\cite{LHCb} using a simplified model of the detector geometry
 and mis-alignment configurations.

%% file: Section__Sample.tex
\section{The simulated event sample}
\label{sec:sample}
A sample of $10^8$ $pp \to \Zmm$ events, at a centre-of-mass-energy of $13$~TeV, is generated with POWHEG Box\cite{powheg,Powheg2}.
These events are subsequently processed with Pythia8~\cite{pythia}, which simulates a QCD parton shower, hadronisation, the underlying event,
and QED final state radiation.
Momentum resolution smearing is considered to be unnecessary for the present study,
which is focused on systematic curvature biases rather than the intrinsic curvature resolution.
Since this work was conducted in the context of precision measurements of electroweak observables with LHCb~\cite{Vesterinen,Pili,Lupton,afb_lhcb,YR_SM_HLLHC}, events are selected with both muons in the pseudorapidity interval $1.7 < \eta < 5$.
It is also required that both muons have $p_T>15$\,GeV and that at least one of the two muons has $p_T>25$\,GeV.
The coordinates are fixed according to a right-handed
Cartesian system with the origin at the $pp$ interaction point. The $x$-axis is oriented horizontally
towards the outside of the LHC ring, the $y$-axis points upwards with respect to the beamline and the
$z$-axis is aligned with the beam direction. The simplified geometry includes a dipole magnetic field along $y$, bending the tracks on the $x-z$ plane and a single measurement plane after the magnet.
Charged particles are deflected along the $x$ axis by $p^{-1} \times 3~\mathrm{GeV}\,\mathrm{m}$, which roughly corresponds to the bending power of the LHCb spectrometer~\cite{LHCb}.
A curvature bias ($\delta$) can be interpreted as a systematic translation of the measurement plane along the $x$ axis:
\begin{equation}
\Delta x = \delta \times 3\,\mathrm{GeV}\,\mathrm{m}.\label{eq:XvsP}
\end{equation}
In Section~\ref{sec:mockup} it is estimated that a $\Delta x$ value of 5\,\mum leads to a bias of $\mathcal{O}$(50)\,MeV 
in the determination of $m_W$ for a single charge. It should be noted that the bias in the measurement of $m_W$ for a single charge is strongly (but not fully) anti-correlated to the corresponding bias in $m_W$ for the opposite charge, leading to partial cancellations when the combination of the two charges is considered. Nevertheless, the precision of the measurement is potentially affected by such biases.
Given that, for example, the LHCb Run-II dataset permits an $\mathcal{O}(10)$\,MeV statistical precision on $m_W$~\cite{Vesterinen}
a simple and reliable method to control this source of bias is required.

%% file: Section__Method.tex
\section{The pseudomass method}
\label{sec:method}
\begin{figure}
    \centerline{
    \includegraphics[width=\figwidth\textwidth]{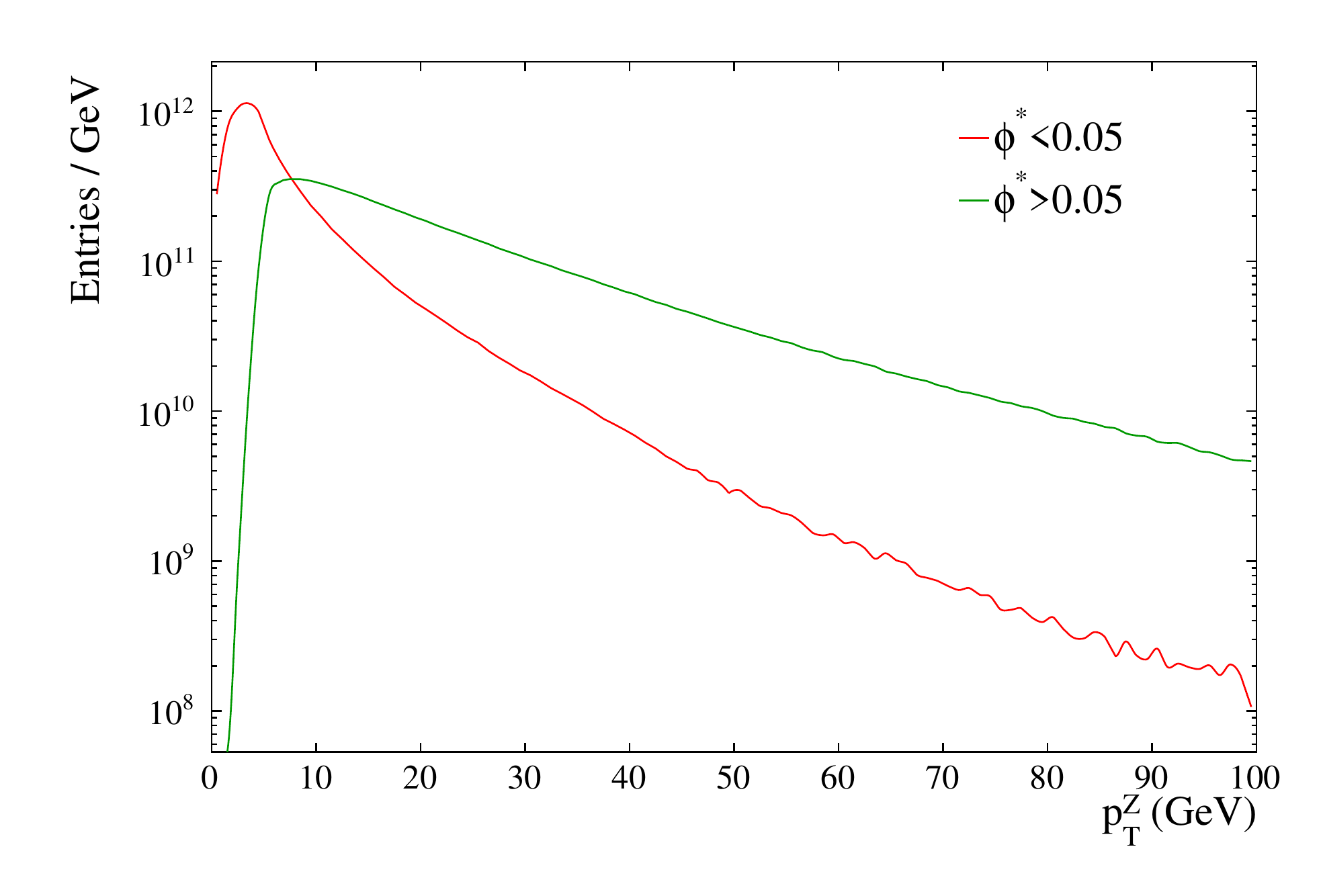}\includegraphics[width=\figwidth\textwidth]{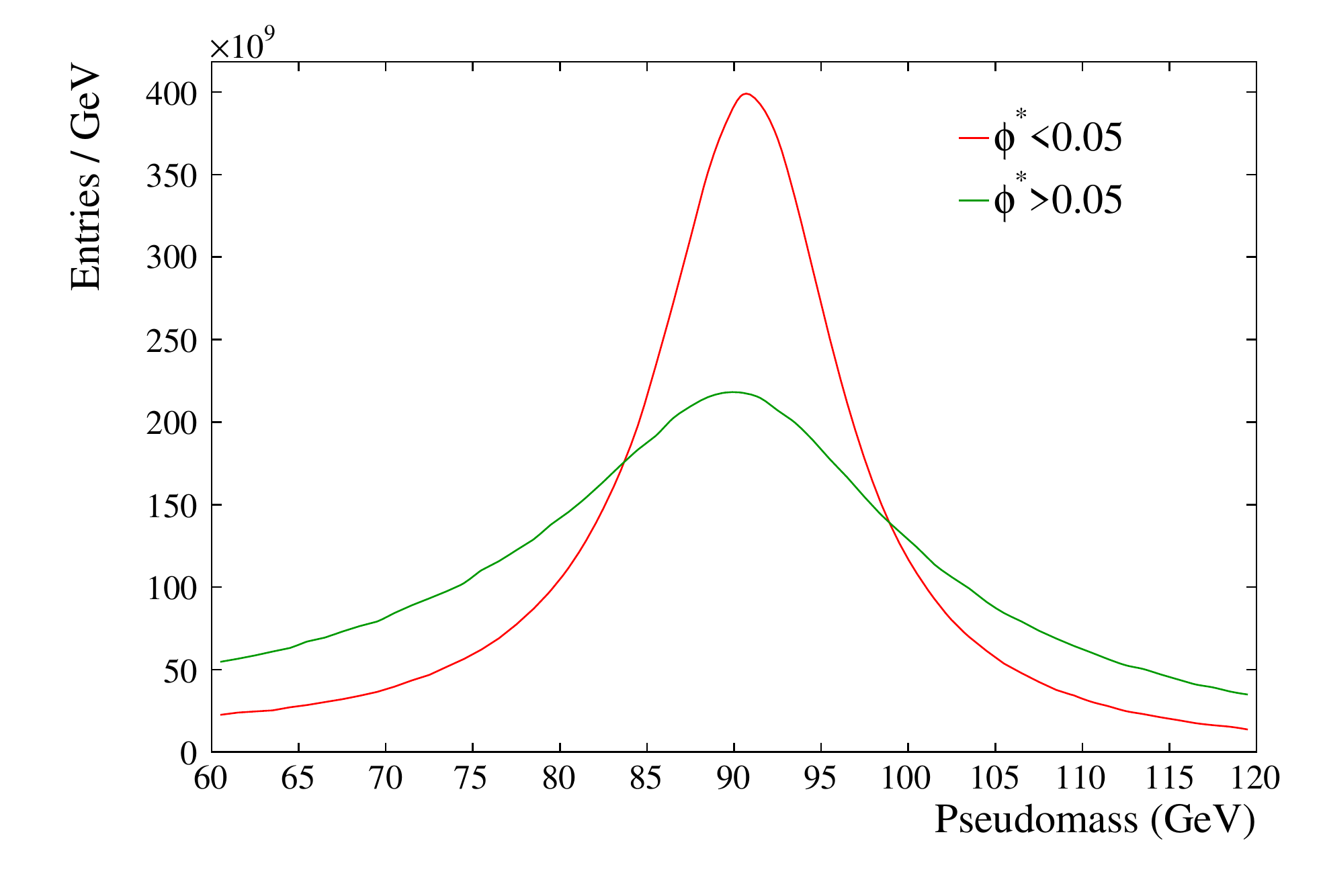}
}
    \caption{The distribution of $p_T^Z$ (left) and the pseudomass (right) for events passing and failing the $\phi^*<0.05$ cut.}\label{fig:phistar}
\end{figure}

\begin{figure}
    \centering
    \includegraphics[width=\figwidth\textwidth]{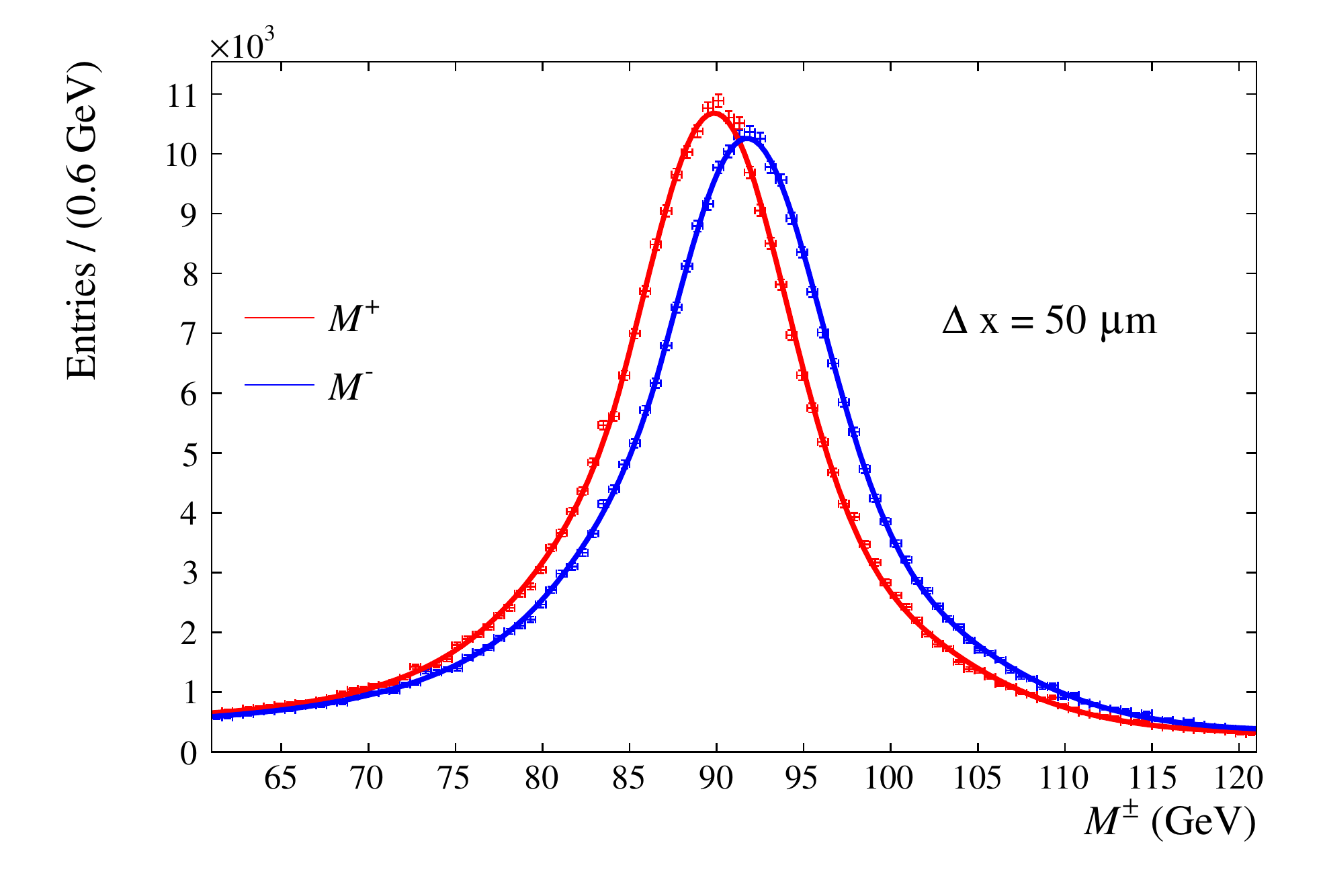}
    \caption{The parametric fit of the $M^+$ and $M^-$ pseudomass distributions, in the presence of
        a $\Delta x = 50$\mum. The signal is modelled
    with a double Crystal Ball function, while the non-peaking component of the $Z/\gamma^*$ line-shape is modelled with an exponential.}\label{fig:pseudomass_fit}
\end{figure}

The proposed method relies on an approximation of the invariant mass of \Zmm decays using the momentum of one muon and only the
\emph{direction} of the other. 
In the context of a measurement of the differential cross section for $p\bar{p} \to \Zmm$ with the D0 experiment, Ref.~\cite{pseudomass} introduced the \emph{pseudomass}\footnote{The definition of
``pseudomass" in D0 is given in terms of leading/sub-leading muons, not positive/negatively charged muons. Furthermore, the pseudomass was used for different purposes, and not for addressing curvature biases.} 
($M^{\pm}$) for each muon charge as:
\begin{equation}
    M^\pm = \sqrt{2p^\pm p_T^\pm \frac{p^\mp}{p_T^\mp}(1-\cos\theta)},\label{eq:pseudomassdef}
\end{equation}
where $p^{\pm}$ and $p_T^{\pm}$ are the momenta and transverse momenta of the $\mu^{\pm}$
and $\theta$ is the opening angle between the two muons. 
The pseudomass is an estimate of the dimuon mass under the assumption that the dimuon system has zero momentum transverse to the bisector of the two lepton transverse momenta\footnote{Ref.~\cite{phistar} refers to this axis as $\hat{b}$.}. 
This assumption is inspired by the fact that most of the \Zmm cross section at hadron colliders is in the region $p_T^Z < m_Z$. A subset of events with smaller $p_T^Z$ values can be selected, independently of the 
muon momenta, using the $\phi^*$ variable~\cite{phistar}, defined as:
\begin{equation}
    \phi^* \equiv  \tan(\phi_{\text{acop}}/2)\sin\theta^*_\eta \approx
    \frac{p_T^Z}{m_Z},\label{eq:phistardef}
\end{equation}
where $\phi_{\text{acop}} = (\pi - \Delta\phi)$ and $\Delta\phi$ is the azimuthal opening angle
between the two leptons and $\cos\theta^*_\eta \equiv
\tanh\bigl (\frac{\eta^- - \eta^+}{2}\bigl )$, and where $\eta^-$ and $\eta^+$ are the pseudorapidities of the
negatively and positively charged leptons, respectively.
A requirement of $\phi^*<0.05$ corresponds to roughly half of the available events.
Figure~\ref{fig:phistar} (left) shows how this requirement selects events with smaller $p_T^Z$ on average,
while Fig.~\ref{fig:phistar} (right) shows that the pseudomass distribution of these events has a narrow width of
$\mathcal{O}$(5)\,GeV. 
Figure~\ref{fig:pseudomass_fit} shows that the peaks of the $M^+$ and $M^-$ distributions are
displaced by around 2~GeV when $\Delta x=$ 50\mum is assumed in the simulation. 
This sensitivity can be exploited to determine the $\delta$ in Eq.~\ref{eq:curvature} through a simultaneous likelihood fit of the $M^+$ and $M^-$ distributions. Both distributions are modelled by a double Crystal Ball (CB) function~\cite{Skwarnicki:1986xj}, plus an
exponential function for the non-peaking component of the $Z/\gamma^*$ line-shape. The mean of the CB function for the $M^{\pm}$ distribution is defined as $\bar{M}(1\pm A)$, where $\bar{M}$ and the asymmetry $A$ are two of the 16 floating parameters of the fit. 
The other floating parameters are the width of the first CB function and the relative width of the second CB (for both $M^+$ and
$M^-$); the $\alpha, n$
parameters describing the tail of the CB; the slope of the exponential component; the normalisation constants of the three shape functions.
The charge-dependent curvature bias can be determined via
\begin{equation}
    \delta \approx A\frac{<\frac{1}{p^+}> + <\frac{1}{p^-}>}{2}.\label{eq:transf}
\end{equation}
Since this approach decouples the effect of curvature biases on the momenta of the $\mu^+$ and $\mu^-$ it is straightforward to determine $\delta$ values in an arbitrary number of bins of $\eta$ and $\phi$.
The curvature biases can be determined independently in each bin.
Unless otherwise specified a binning scheme with 8 (12) $\eta$ ($\phi$) bins is used in the remainder of this paper.

It is anticipated that the determination of the $\delta$ values with Eq.~\ref{eq:transf} will be slightly biased
by the forward-backward asymmetry ($A_{FB}$) in the \Zmm process, which implies a small difference in the kinematic distributions of the $\mu^+$ and $\mu^-$.
However, Eq.~\ref{eq:transf} relies on the asymmetry between the
pseudomass peak positions and  $A_{FB}$ is minimal for dimuon masses close to the peak of the $Z$ resonance.
This bias is nevertheless evaluated and addressed in the following section.

%% file: Section__Validation.tex
\section{Validation of the method}
\label{sec:validation}
The pseudomass method is validated with 30 toy experiments using statistically independent subsets of the 
\Zmm event sample described in Section~\ref{sec:sample}. 
When no mis-alignment is simulated the method should return $\Delta x$ values that are statistically compatible with zero.

\begin{figure}
    \centering
    \includegraphics[width=1.0\textwidth]{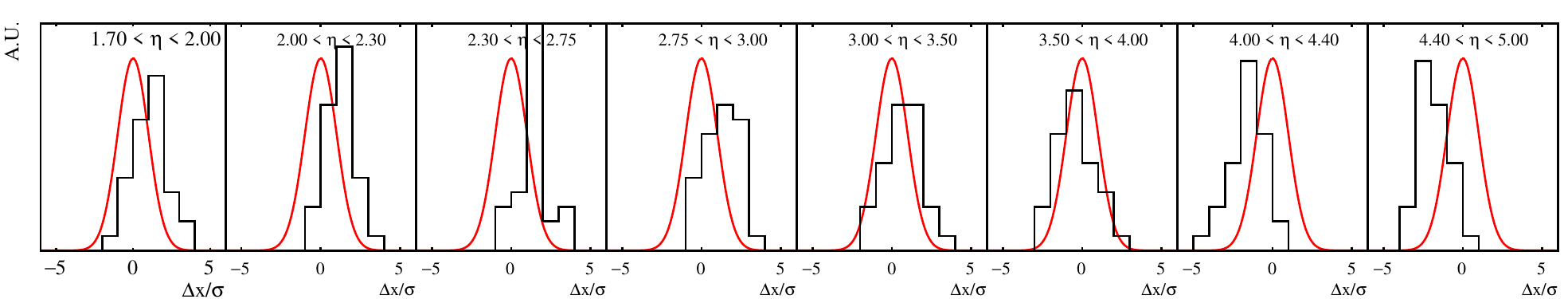}\\
    \includegraphics[width=1.0\textwidth]{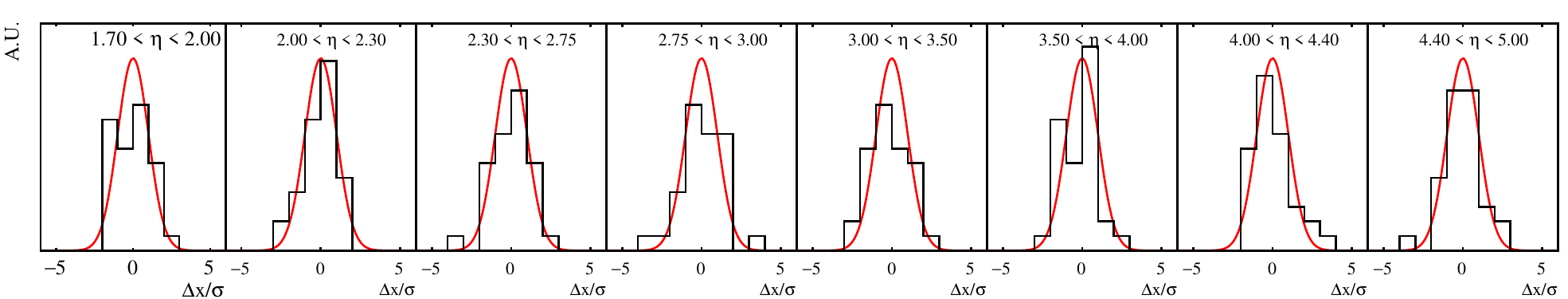}\\
    \includegraphics[width=1.0\textwidth]{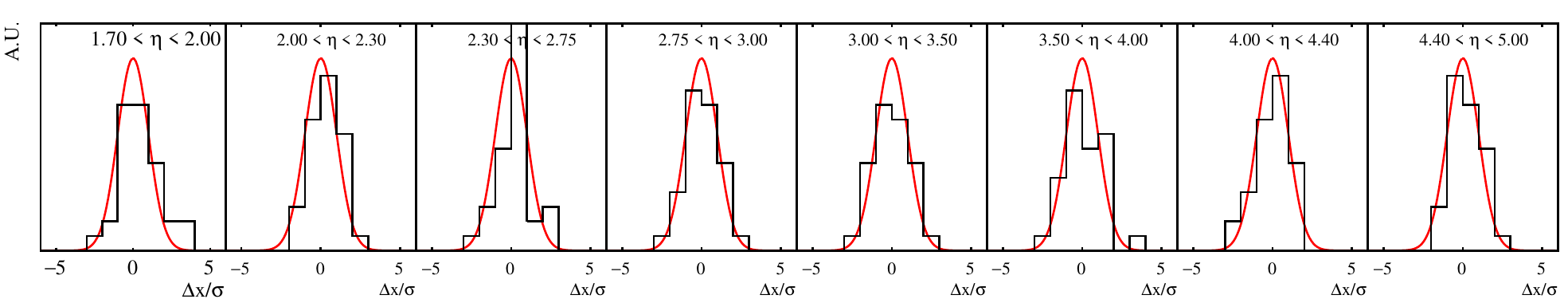}
    \caption{The distribution of the difference between the measured $\Delta x$ values
        calculated with the pseudomass method and the
    expectation ($\Delta x=0$\mum) divided by the uncertainty of the measured parameter in bins of
    $\eta$ for 30 toy experiments. The upper two rows correspond to Eq.~\ref{eq:transf}.
    In the central row the muon charges have been randomised. In the lower row Eq.~\ref{eq:transf_corr} is used.}\label{fig:pulls_removeBias}
\end{figure}

\begin{figure}
    \centerline{
    \includegraphics[width=\figwidth\textwidth]{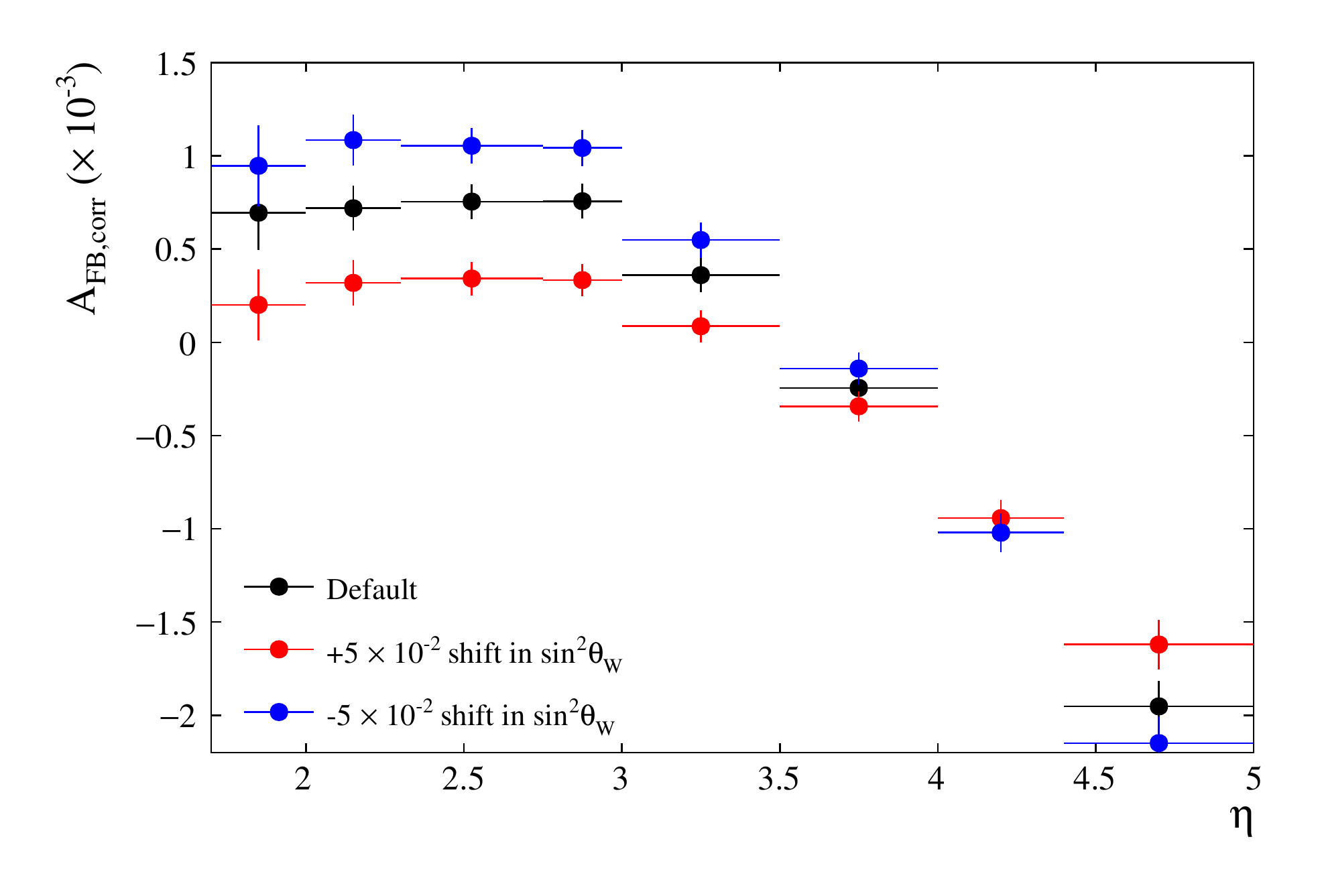}\includegraphics[width=\figwidth\textwidth]{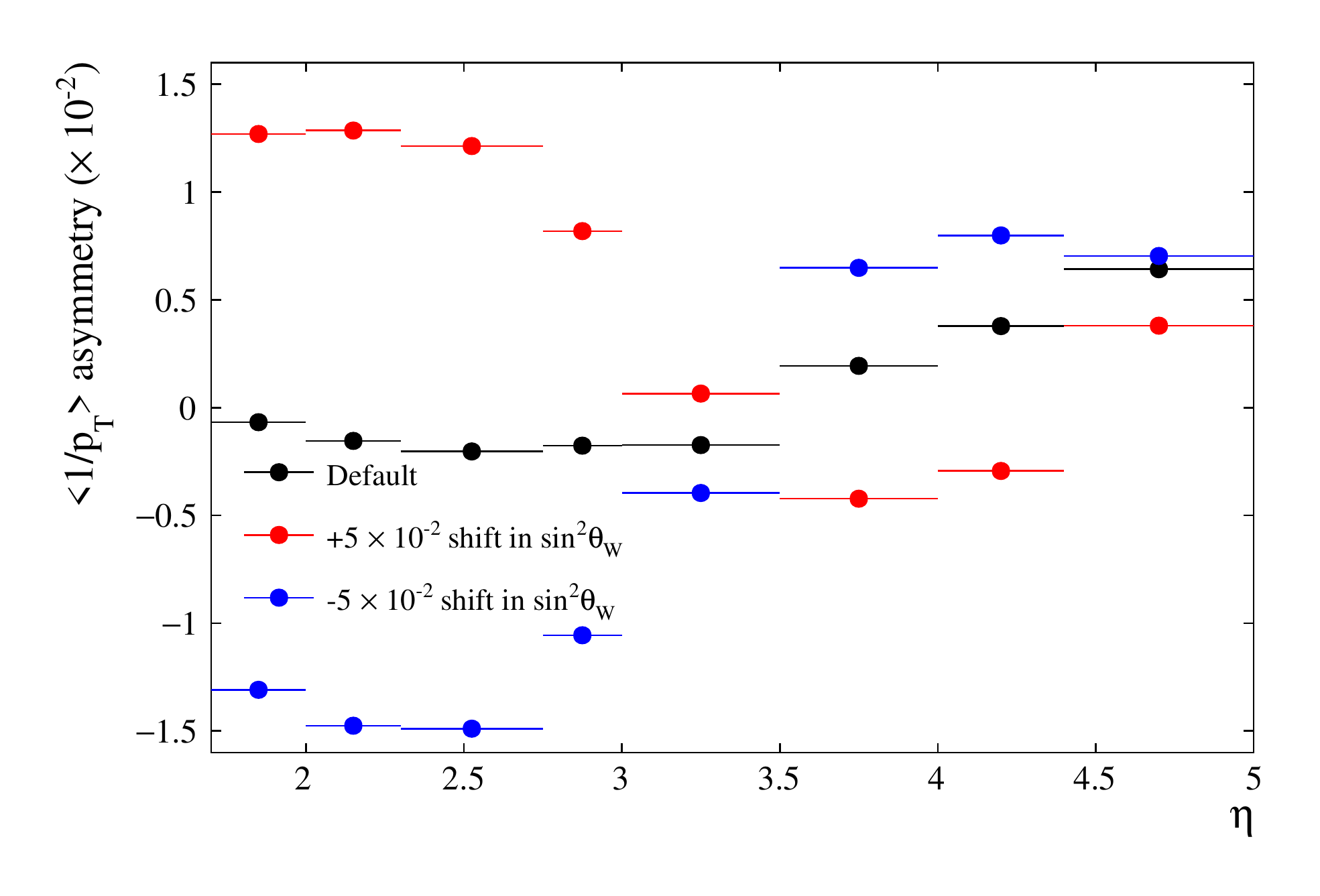}}
    \caption{The pseudomass (left) asymmetry and muon 1/$p_T$ (right) asymmetry in bins of the muon pseudorapidity.
      The asymmetries are shown for three different values of \stw, including two extreme variations (red and blue) around the nominal value (black).}\label{fig:correctionBias}
\end{figure}

Figure~\ref{fig:pulls_removeBias} (upper row) shows the distribution over the 30 toys of the \emph{pull}, i.e. the difference between the
correction derived using the proposed method and the expectation ($\Delta x=0$\mum), 
divided by the statistical uncertainty of the measured
parameter in each of the $\eta$ bins.
In this study an integration over $\phi$ bins is performed.
Given the forward-backward asymmetry it is not surprising to see that the pull
distributions have means that are systematically shifted 
from zero, by up to three standard deviations, with a strong dependence on $\eta$.
Fig.~\ref{fig:pulls_removeBias} (central row) shows that this bias is eliminated when the charges of
the $Z$ boson decay products are randomised, so
that the number of $\mu^+$ and $\mu^-$ falling in each $[\eta,\phi]$ bin is the same (i.e.
effectively ``switching off" the forward-backward asymmetry). 
The small bias can be corrected using simulation but,
given the intended application of the method to measurements of $m_W$ and \stw, careful attention is required for the dependence of this bias on the value of \stw assumed in the simulation.
A larger (than in the 30 toy experiments) sample of around 10$^7$ events is
used to determine this correction with high statistical precision.
Figure~\ref{fig:correctionBias} (left) shows the resulting pseudomass asymmetry values in bins of $\eta$
for the nominal value of \stw (black points) and $\pm 5 \times 10^{-2}$ variations (red and blue points).
It is reassuring to see that the pseudomass asymmetry shifts by at most $5 \times 10^{-4}$ for these extreme
variations, corresponding to roughly $\pm 300$ times the uncertainty on the current world average value of \stw~\cite{PDG}.
Figure~\ref{fig:correctionBias} (right) shows that the same variations in \stw lead to far greater changes in the profile of the mean $1/p_T$ asymmetry versus $\eta$.
The pseudomass asymmetry therefore appears to be better suited than the $1/p_T$ asymmetry for the determination
of curvature biases with \Zmm events.
A small correction for this bias can now be included in Eq.~\ref{eq:transf}, as follows:
\begin{equation}
    \delta \approx (A - A_{FB,\text{corr}})\frac{<\frac{1}{p^+}> +
            <\frac{1}{p^-}>}{2},\label{eq:transf_corr}
\end{equation}
where $A_{FB,\text{corr}}$ corresponds to the black points in Fig.~\ref{fig:correctionBias}.
Figure~\ref{fig:pulls_removeBias} (lower row) shows that using Eq.~\ref{eq:transf_corr} the curvature biases
can be determined without bias, since the pull distributions of the $\Delta x$ values are consistent with standard normal distributions
in all the bins considered.

%% file: Section__Application.tex
\section{Application of the method}
\label{sec:application}
The pseudomass method is tested on a subset of $3 \times 10^{6}$ events from the sample described in
Section~\ref{sec:sample}. 
The muon momenta are re-calculated to mimic the effects of five mis-alignment scenarios
that are representative of the LHCb detector~\cite{LHCBperf}.
There is a distinction between {\em coherent} translations/rotations of the entire measurement plane
and {\em incoherent} translations in bins of $\eta$ and $\phi$.
In all cases the \emph{additional} deflection ($\Delta x$) of a track along $x$ due to the introduced mis-alignment is calculated, and the corresponding variation in the momentum is derived from Eq.~\ref{eq:XvsP}. 
The five mis-alignment scenarios are configured as follows.
\begin{enumerate}
    \item Coherent translation along $x$ ($\Delta x$ = 50\mum).
    \item Coherent translation along $z$ ($\Delta z$ = 100\mum): the corresponding deflection along $x$ is
\begin{equation}
    \Delta x = \Delta z \frac{\cos\phi}{\sinh\eta}.
\end{equation}
\item Coherent rotation in the $x-y$ plane ($R_z$ = 0.2\,mrad): the corresponding deflection along $x$ is
\begin{equation}
\Delta x = \frac{\cos(\phi + R_z) - \cos\phi}{\cos\phi}\frac{\text{m}}{p/\text{GeV}}.
\end{equation}
\item Incoherent translation along $x$, with the $\Delta x$ values randomly sampled from a symmetric Gaussian
  distribution with a width of 100\mum in each of the $\eta$ and $\phi$ bins corresponding to the
  same binning scheme as the pseudomass corrections.
\item Incoherent translation along $x$ with five $\eta$ bins instead of eight,
  so that the binning scheme is slightly different to that used in the pseudomass corrections.
\end{enumerate}

\begin{figure}
    \centerline{
    \includegraphics[width=\figwidth\textwidth]{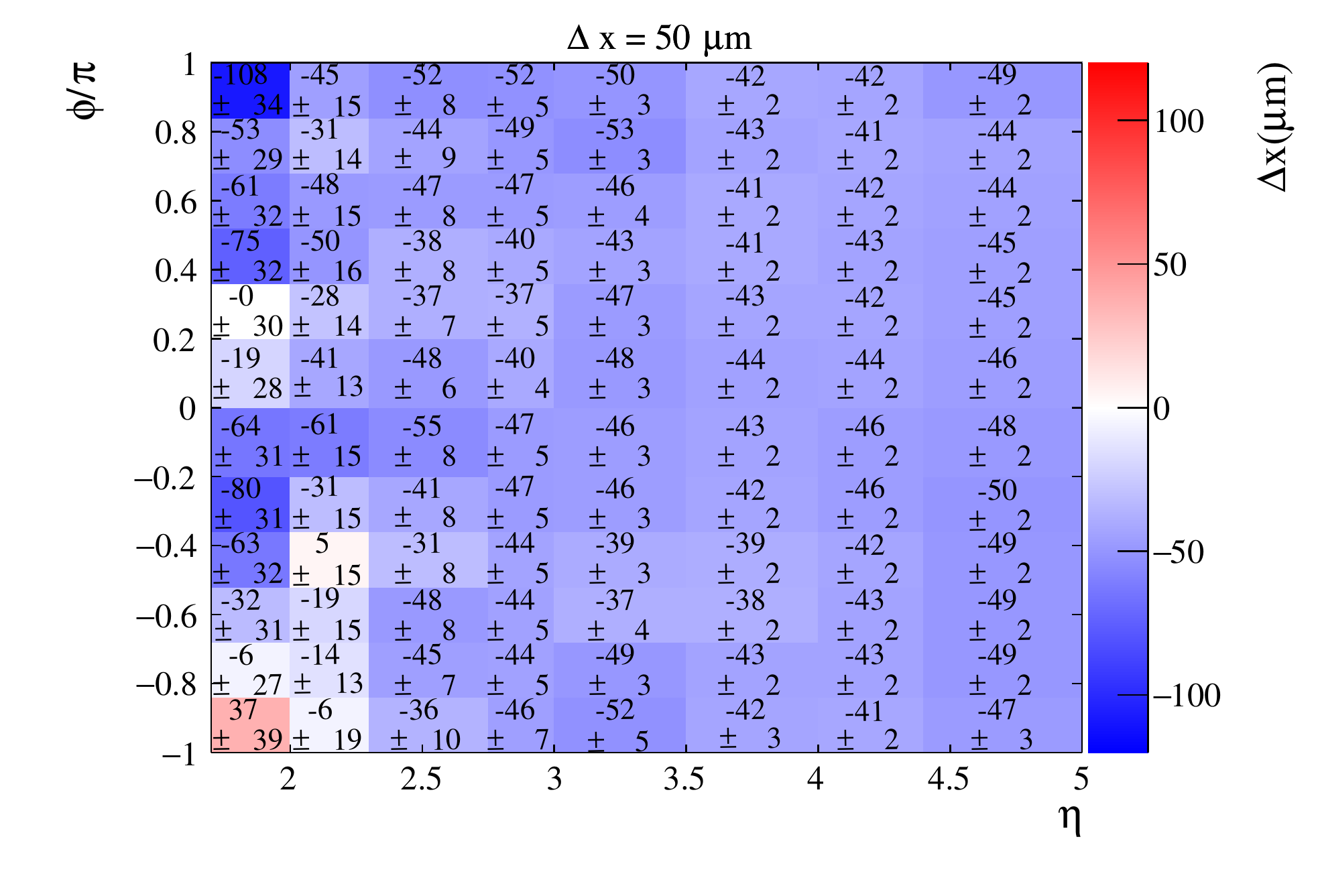}
    \includegraphics[width=\figwidth\textwidth]{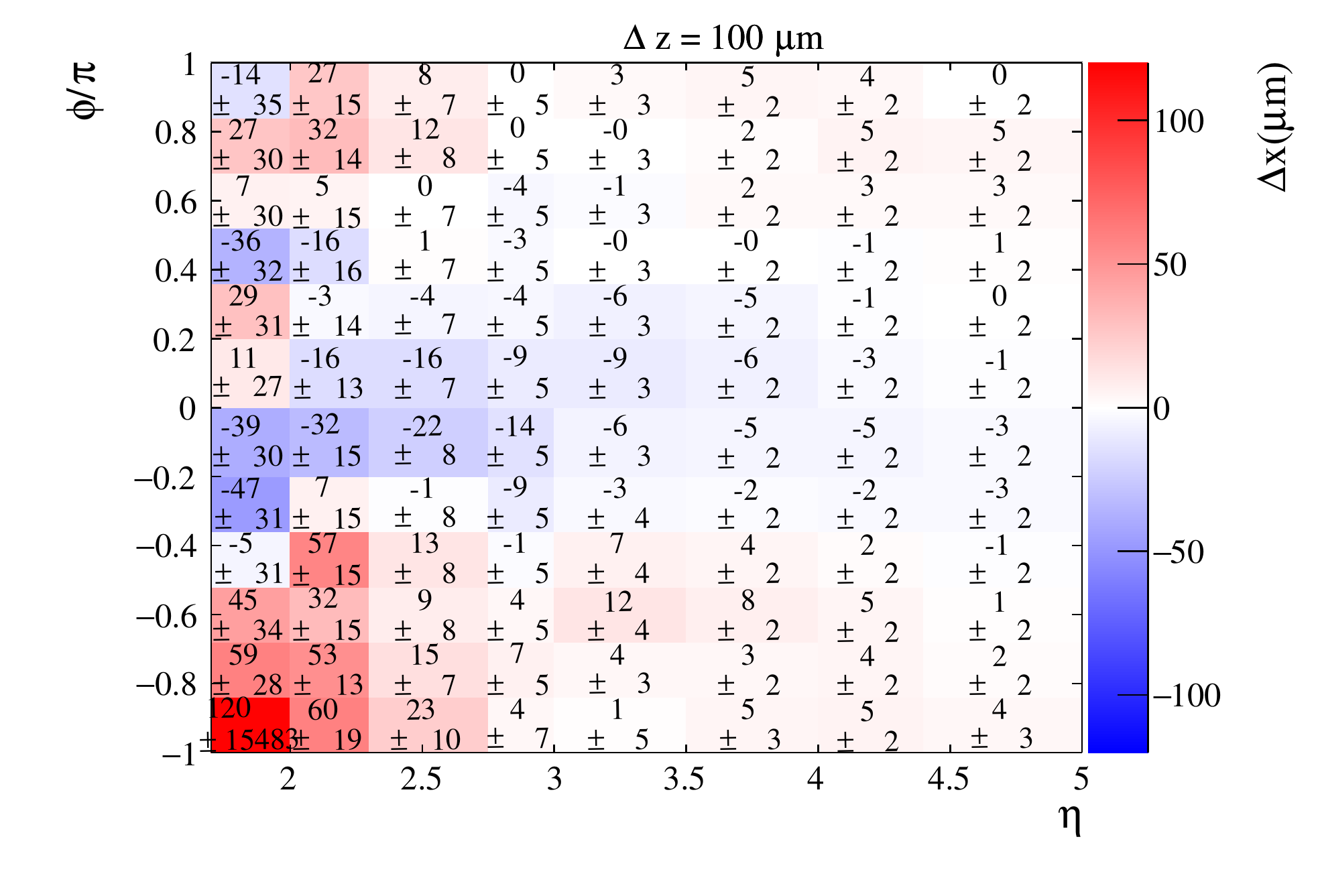}}
    \centerline{
     \includegraphics[width=\figwidth\textwidth]{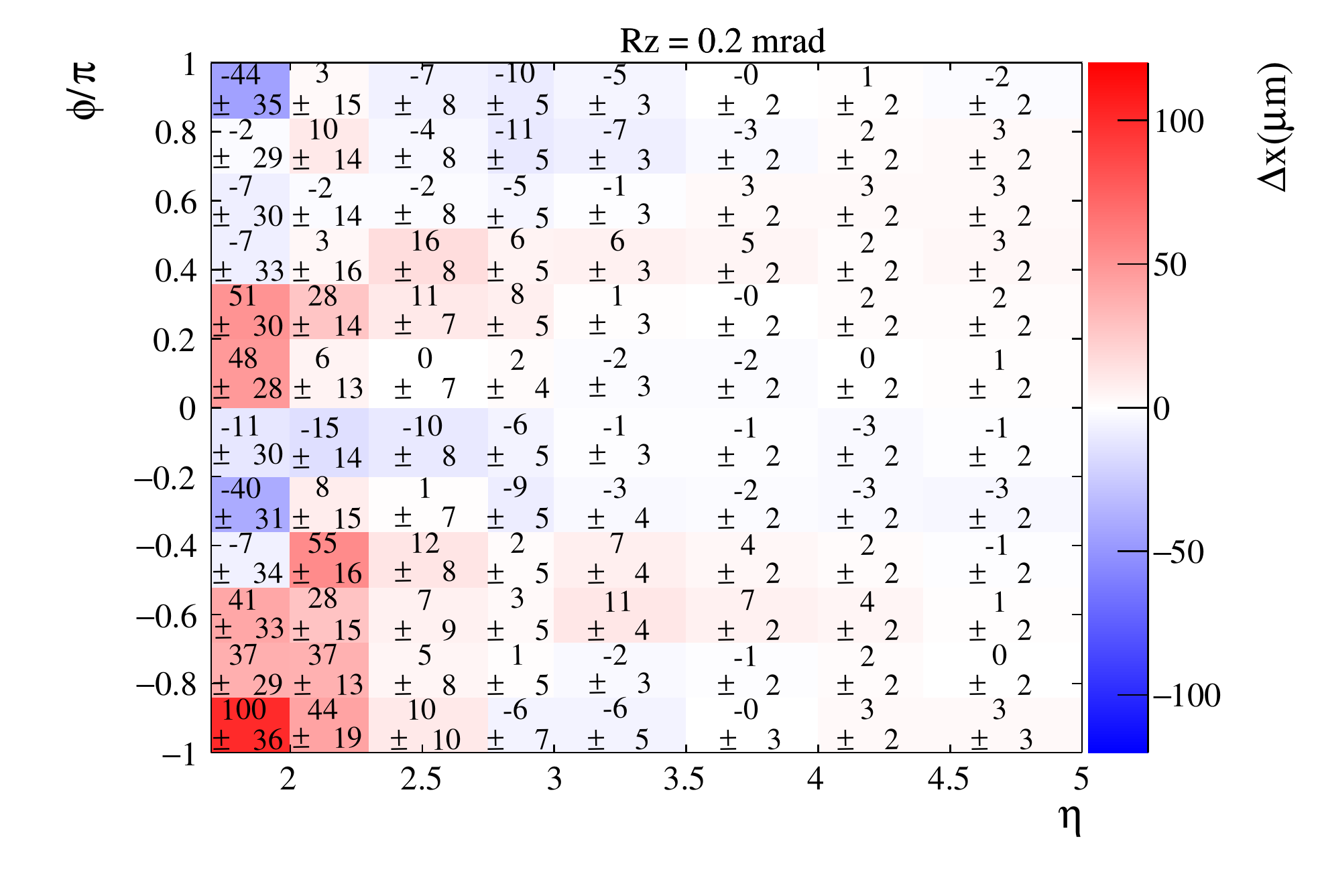}
     \includegraphics[width=\figwidth\textwidth]{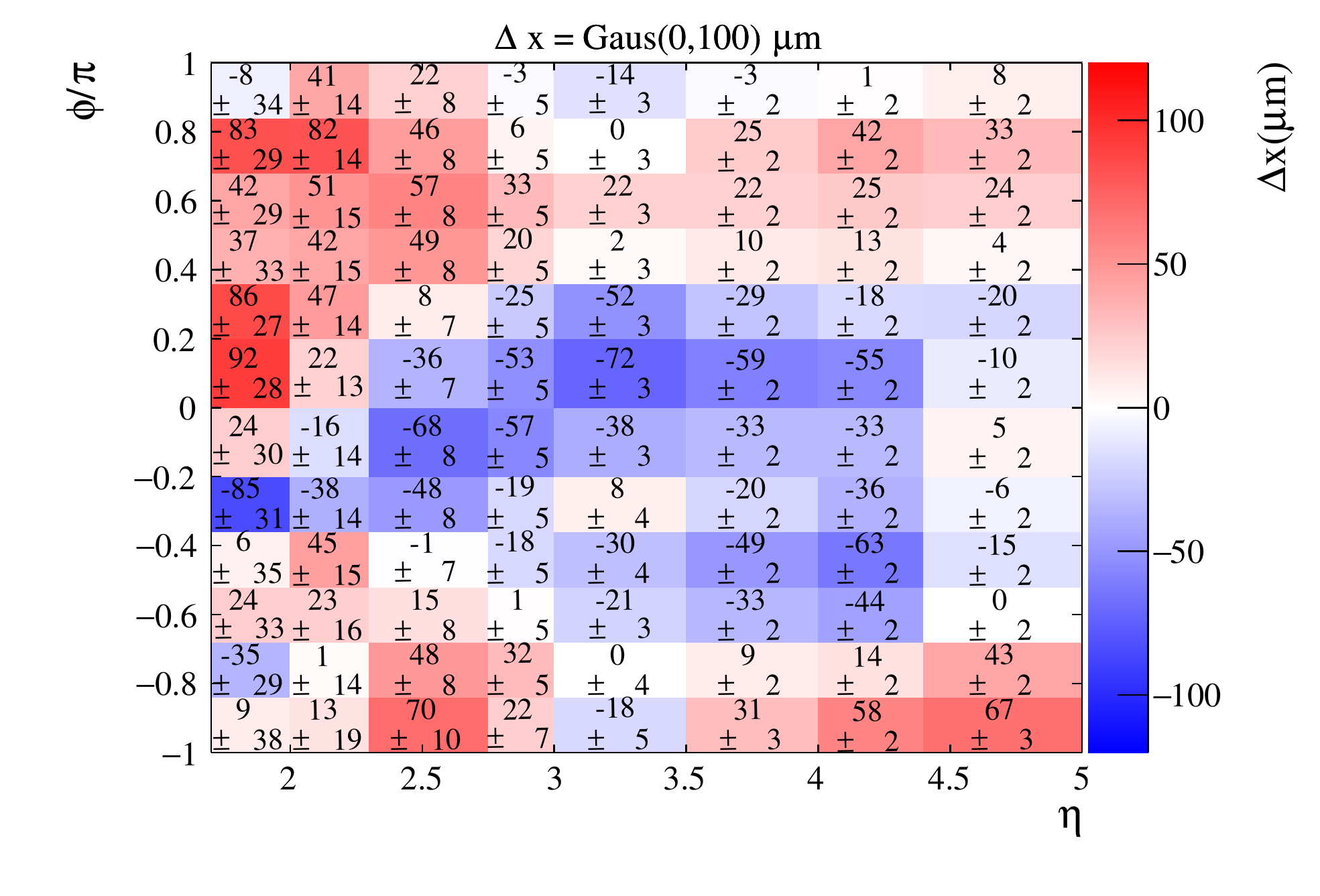}}
 \caption{The pseudomass correction in bins of [$\eta,\phi$] for (clockwise from upper left) a coherent
 $\Delta x$ = 50\mum, coherent $\Delta z = 100$\mum, incoherent $\Delta x = \text{Gaus}(0,100)$\mum, coherent $R_z = 0.2$\,mrad.}\label{fig:maps}
\end{figure}

\begin{figure}
    \centerline{
\includegraphics[width=\figwidth\textwidth]{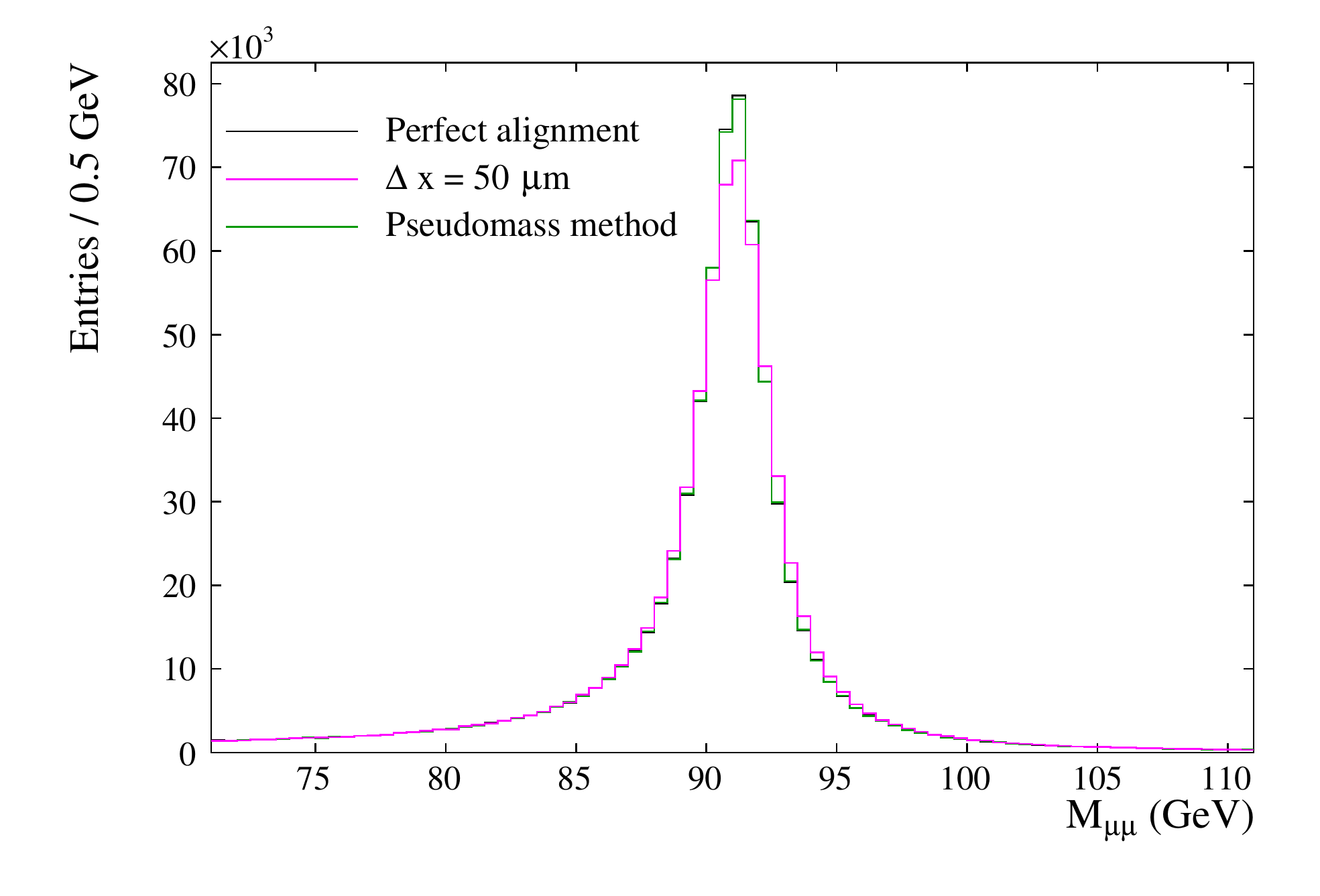}\includegraphics[width=\figwidth\textwidth]{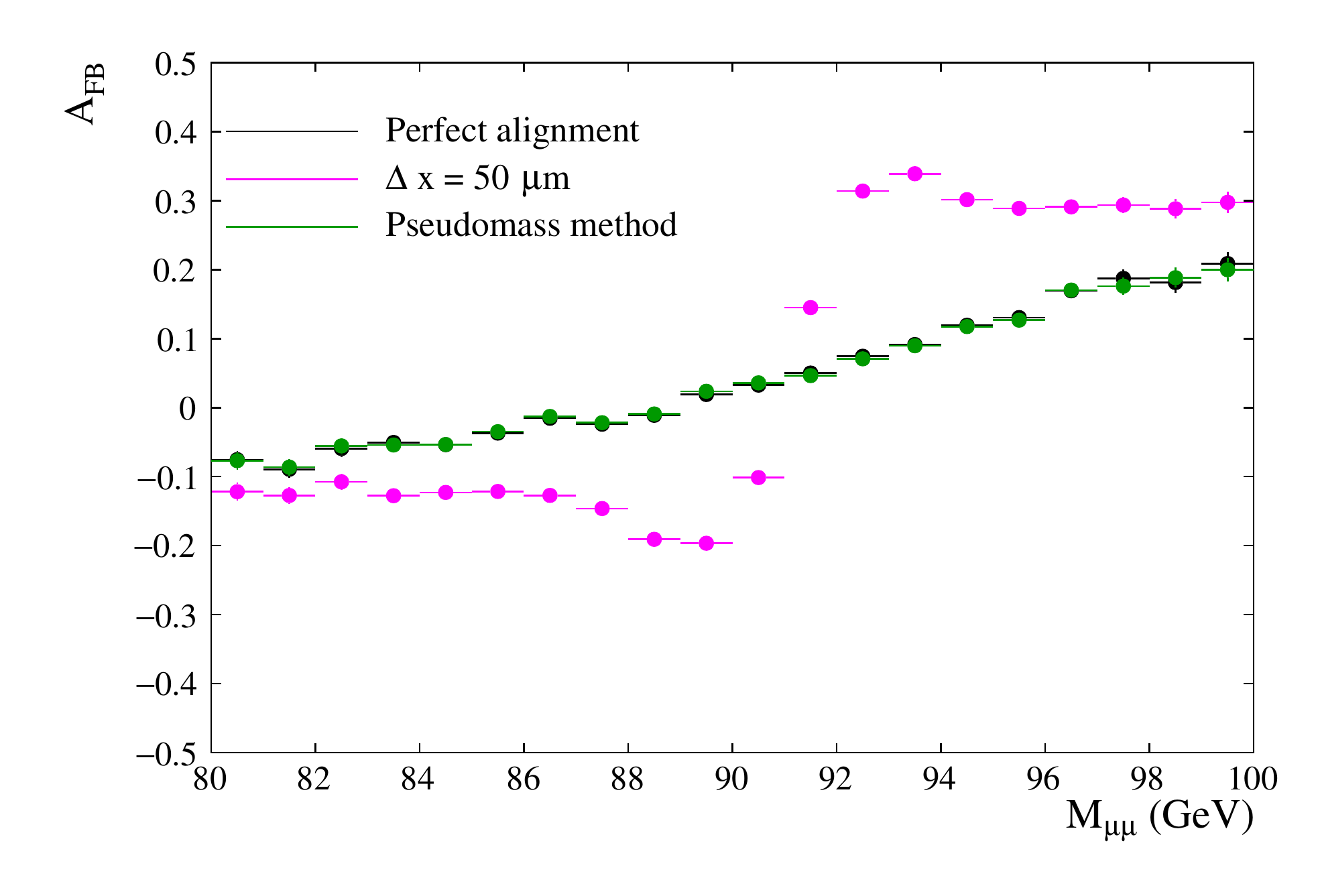}}
\caption{The dimuon invariant mass distribution (left) and the forward backward asymmetry in bins
    of the dimuon invariant mass (right), for simulated events with a coherent $\Delta x$ =
50\mum (magenta), after applying the pseudomass corrections (green), and
the unbiased distribution, with no mis-alignment applied
(black).}\label{fig:Zmass_globTx}
\end{figure}

\begin{figure}
    \centerline{  
    \includegraphics[width=\figwidth\textwidth]{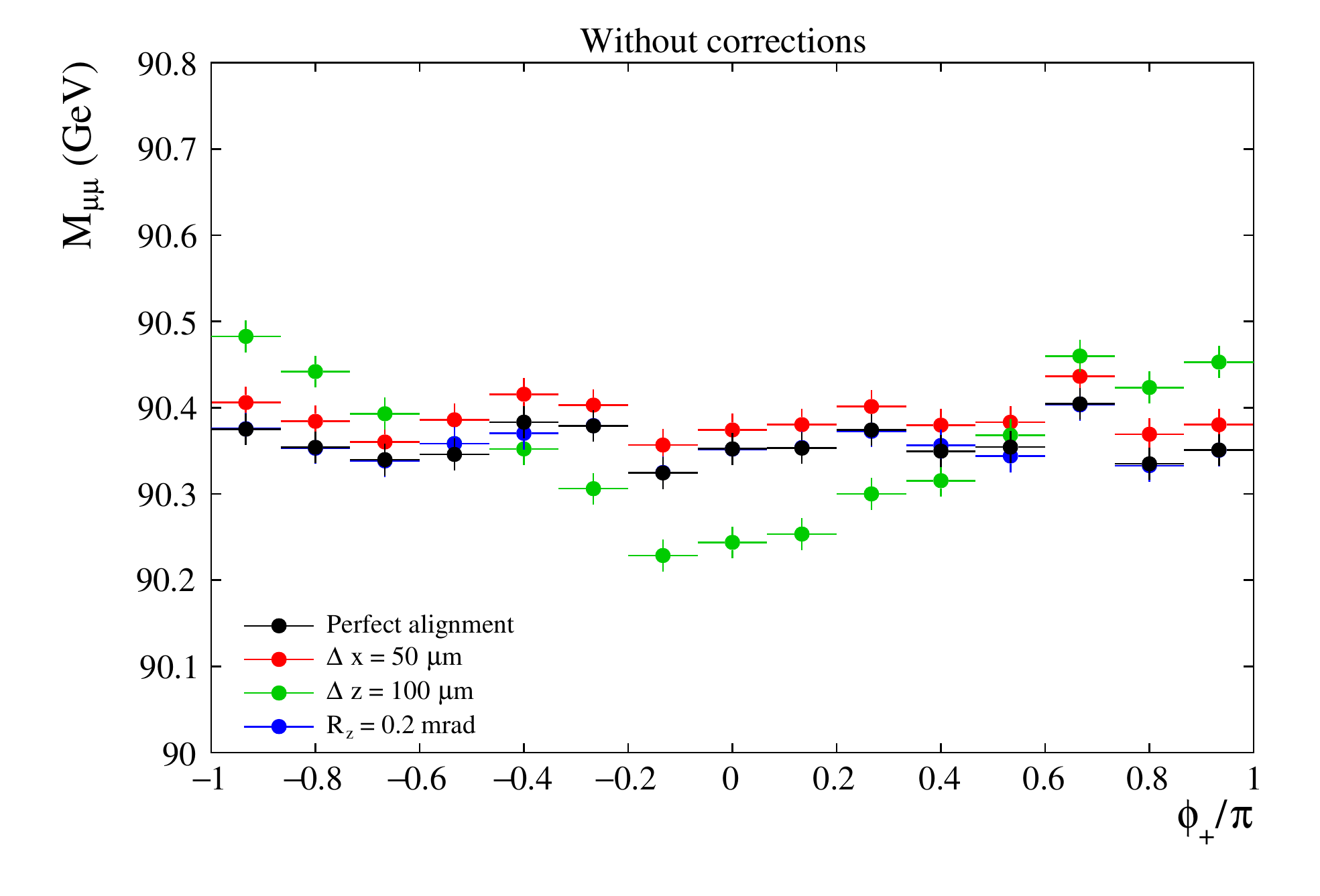}\includegraphics[width=\figwidth\textwidth]{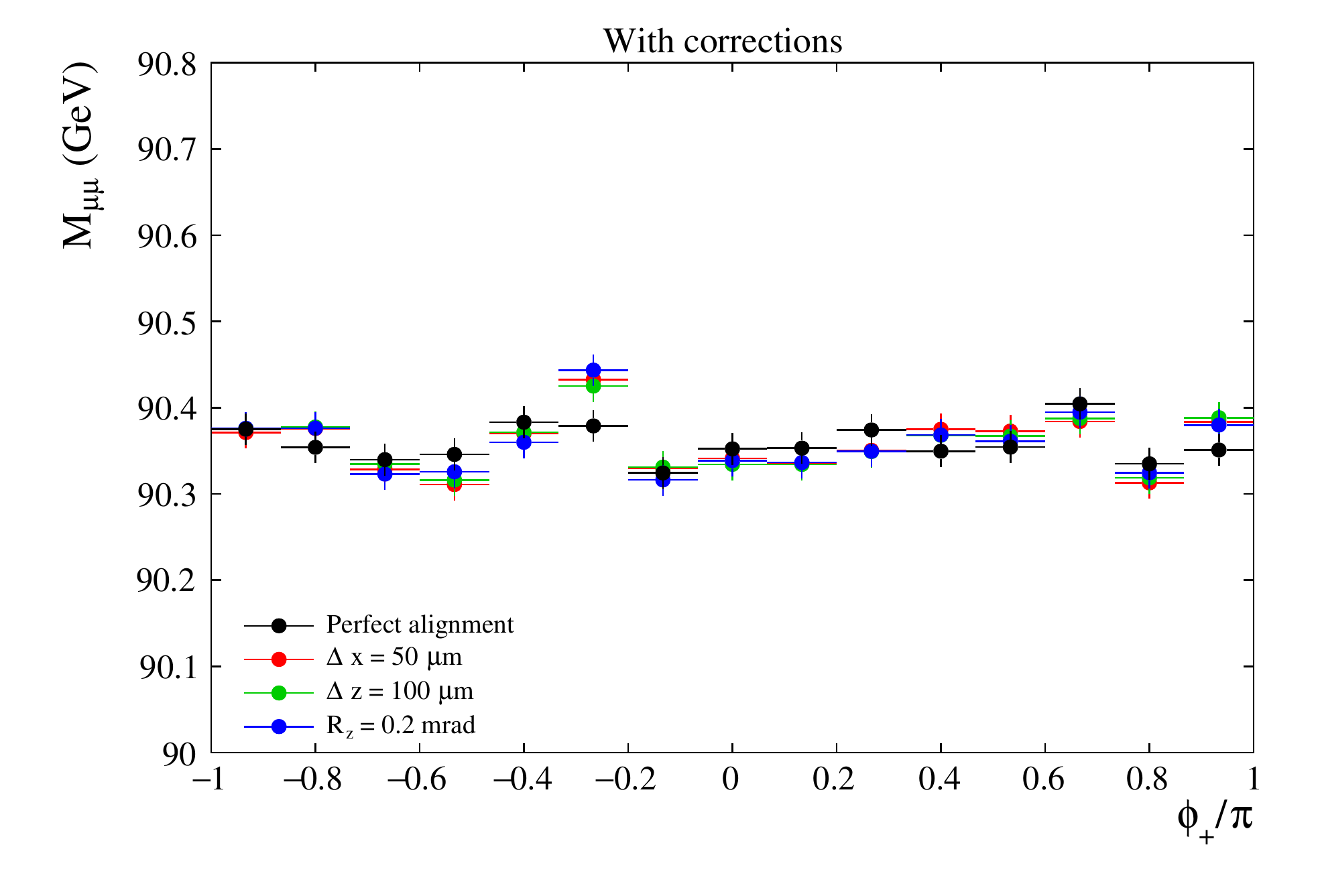}}
        \centerline{  
    \includegraphics[width=\figwidth\textwidth]{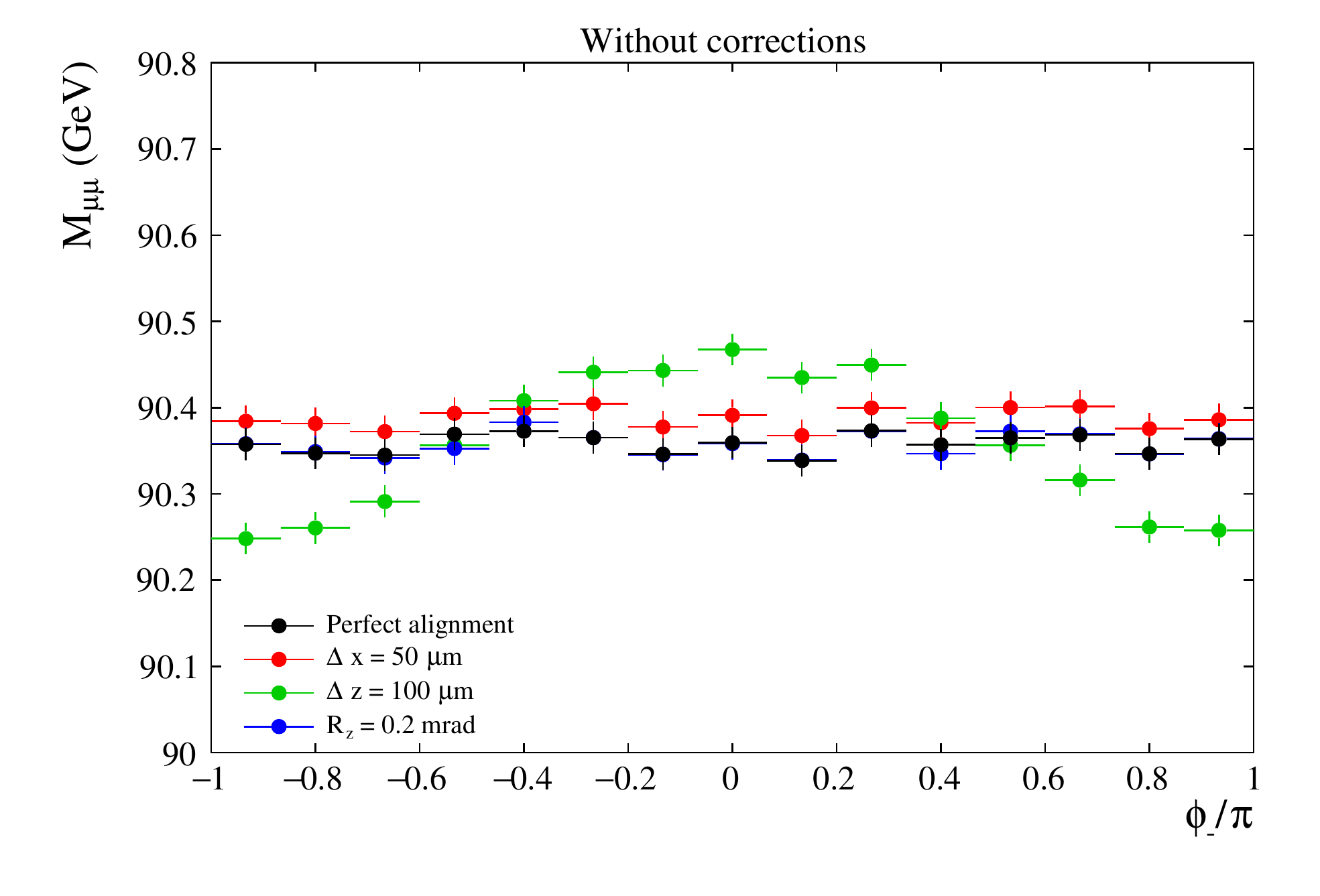}\includegraphics[width=\figwidth\textwidth]{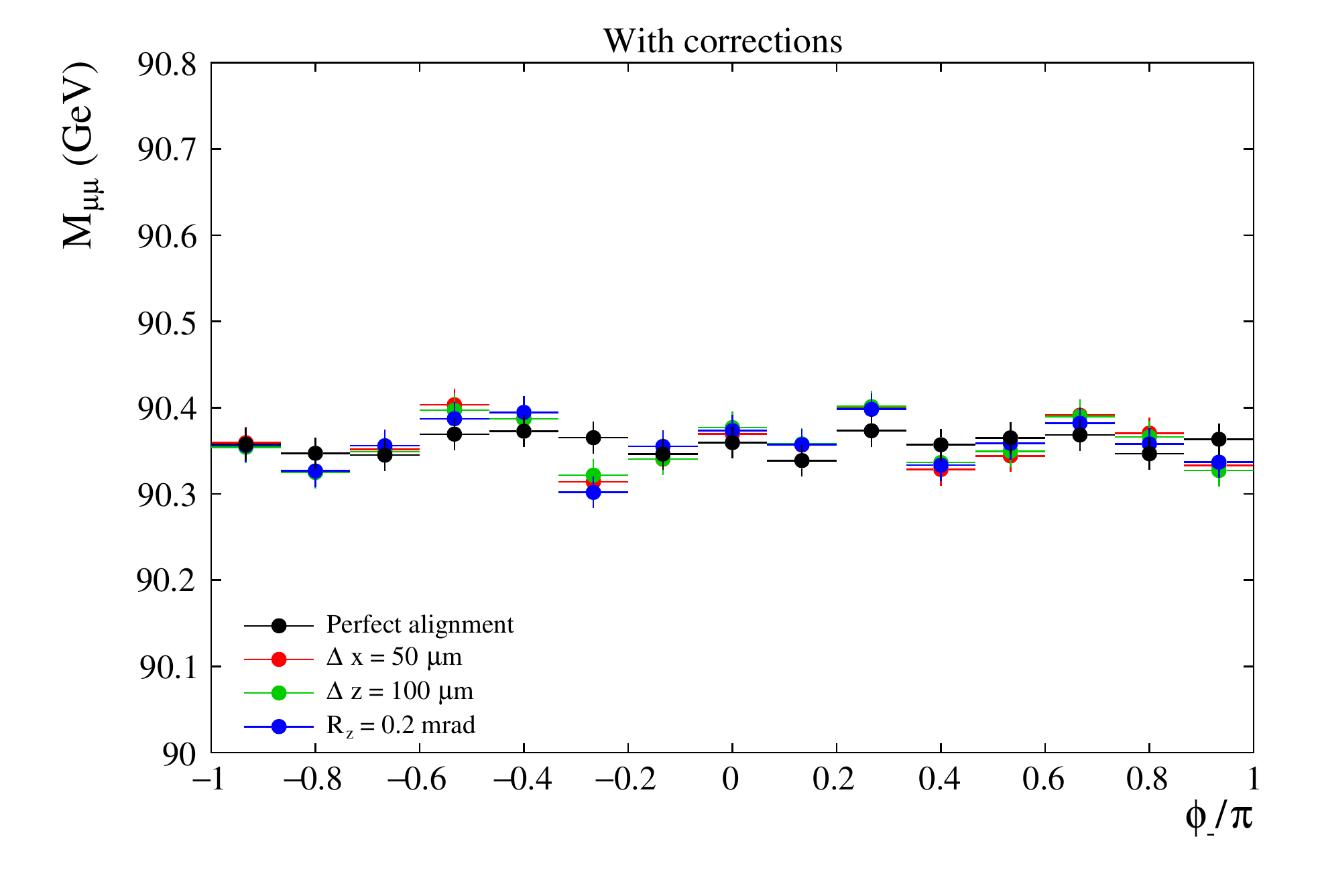}}
        \centerline{  
    \includegraphics[width=\figwidth\textwidth]{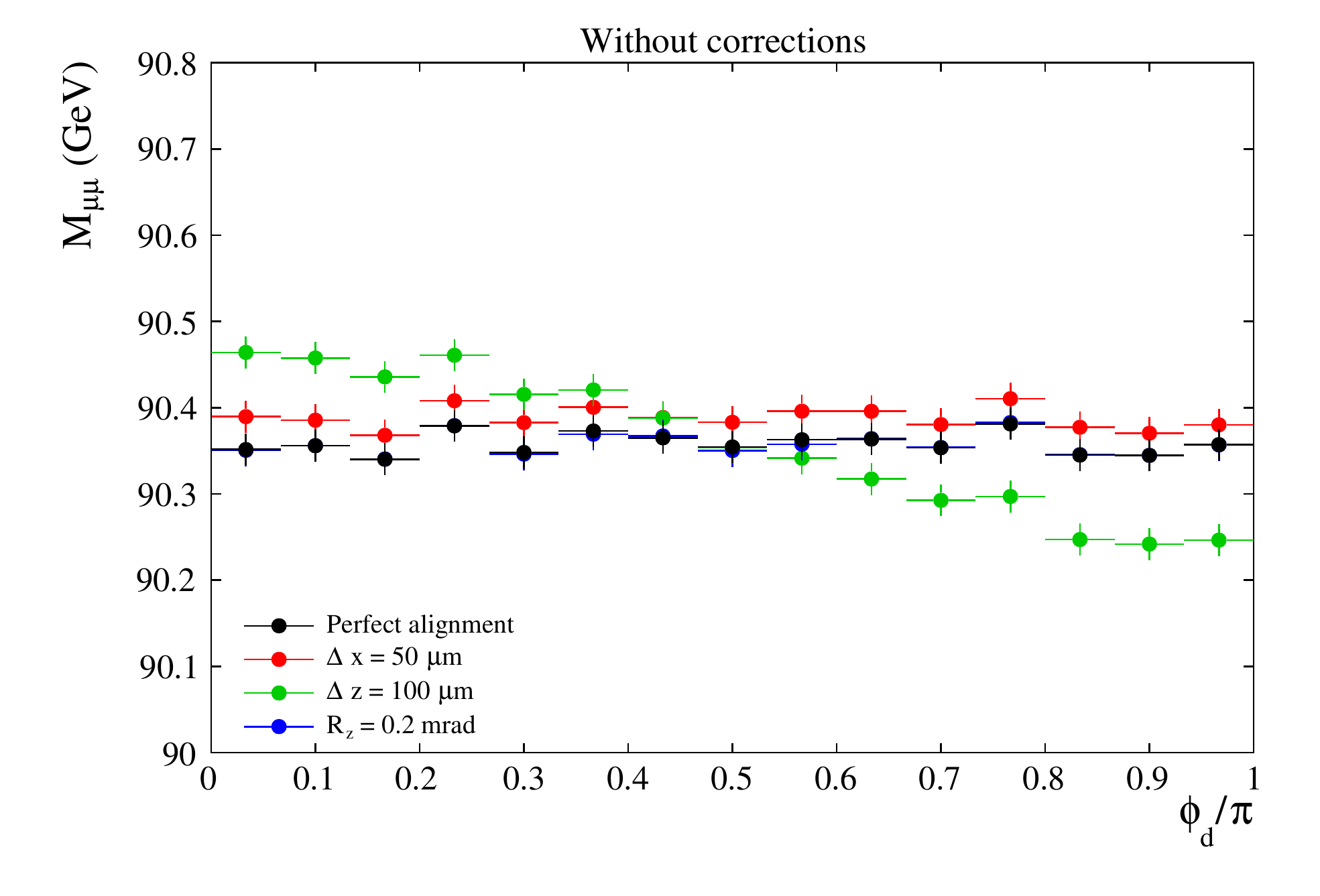}\includegraphics[width=\figwidth\textwidth]{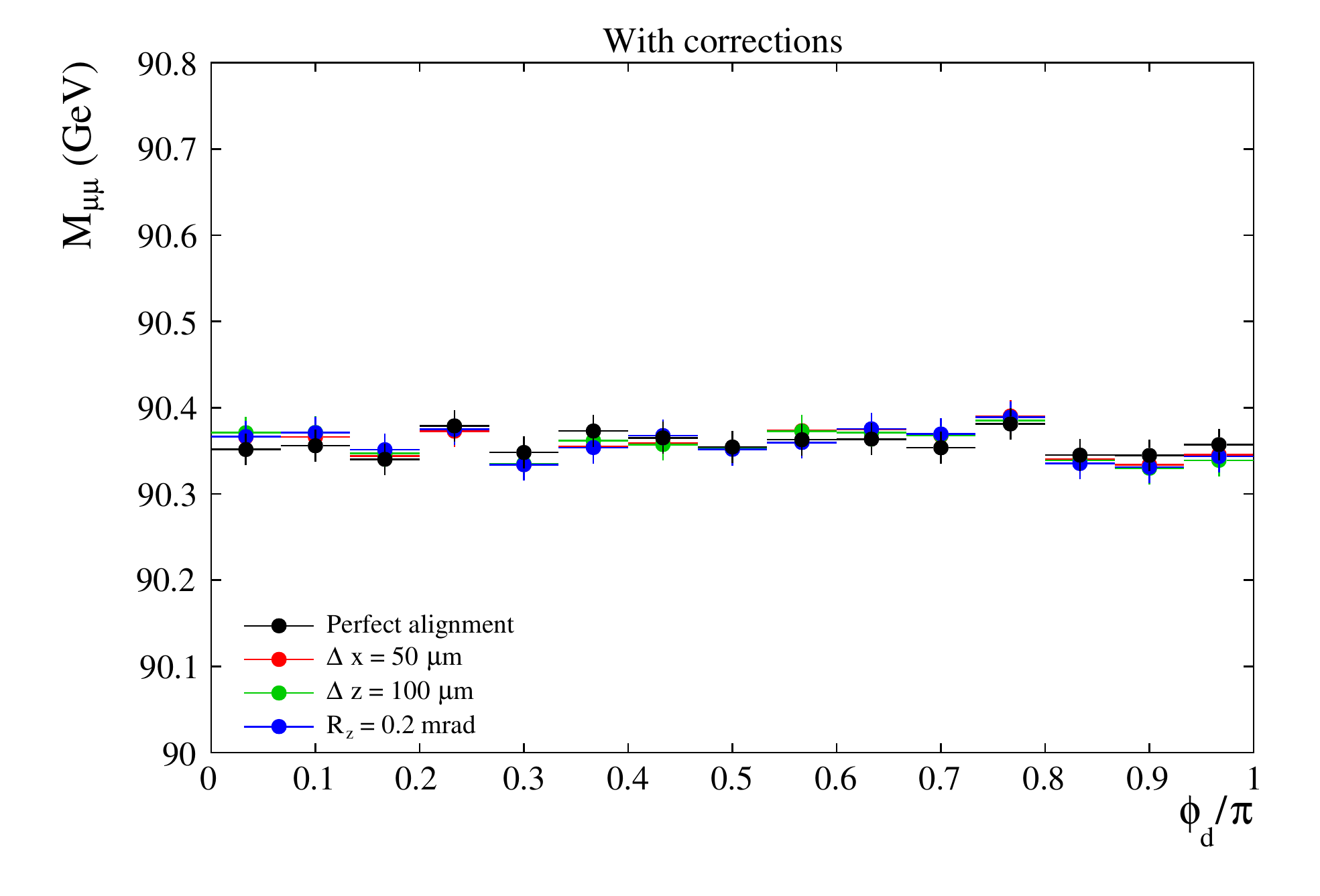}}
        \centerline{  
    \includegraphics[width=\figwidth\textwidth]{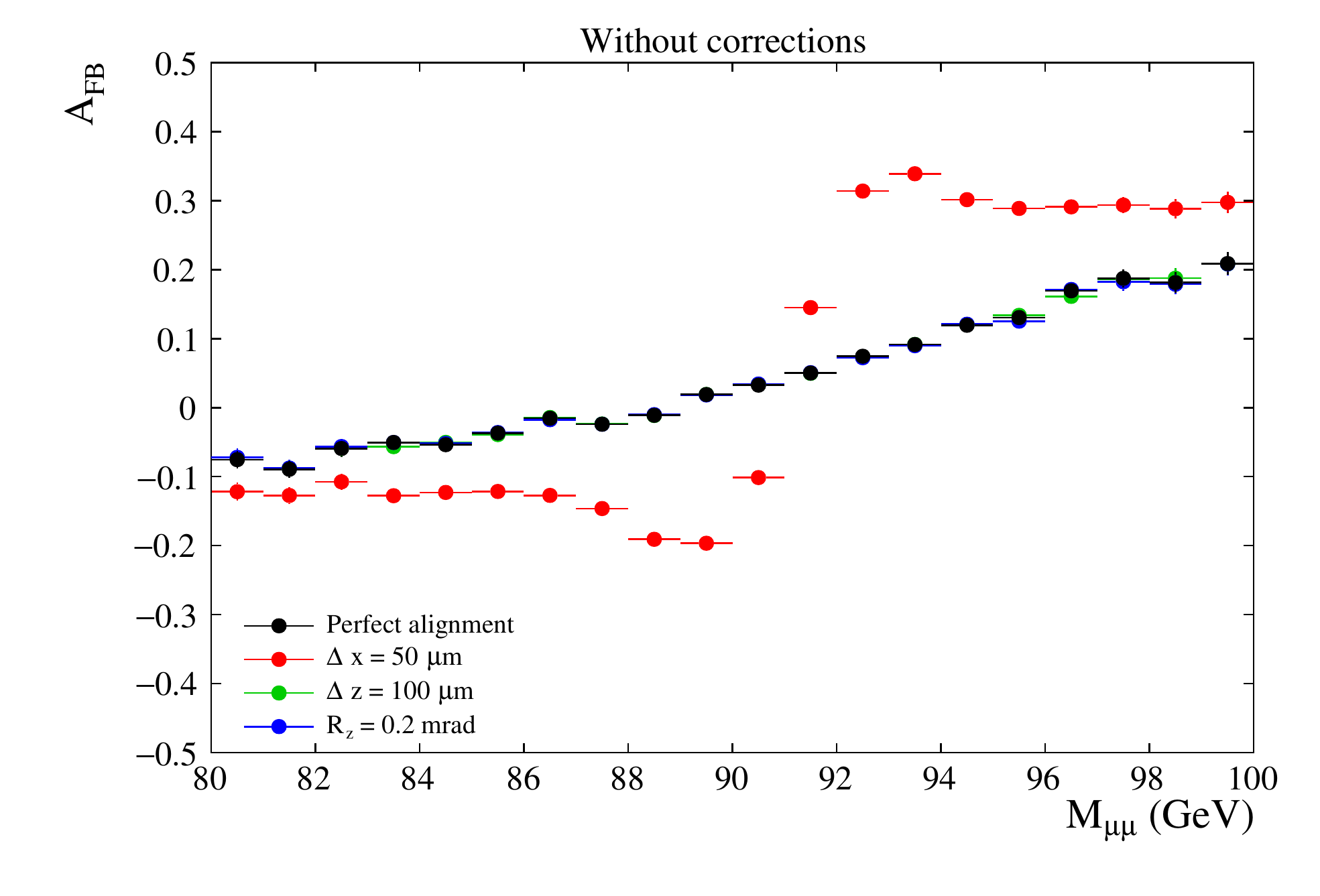}\includegraphics[width=\figwidth\textwidth]{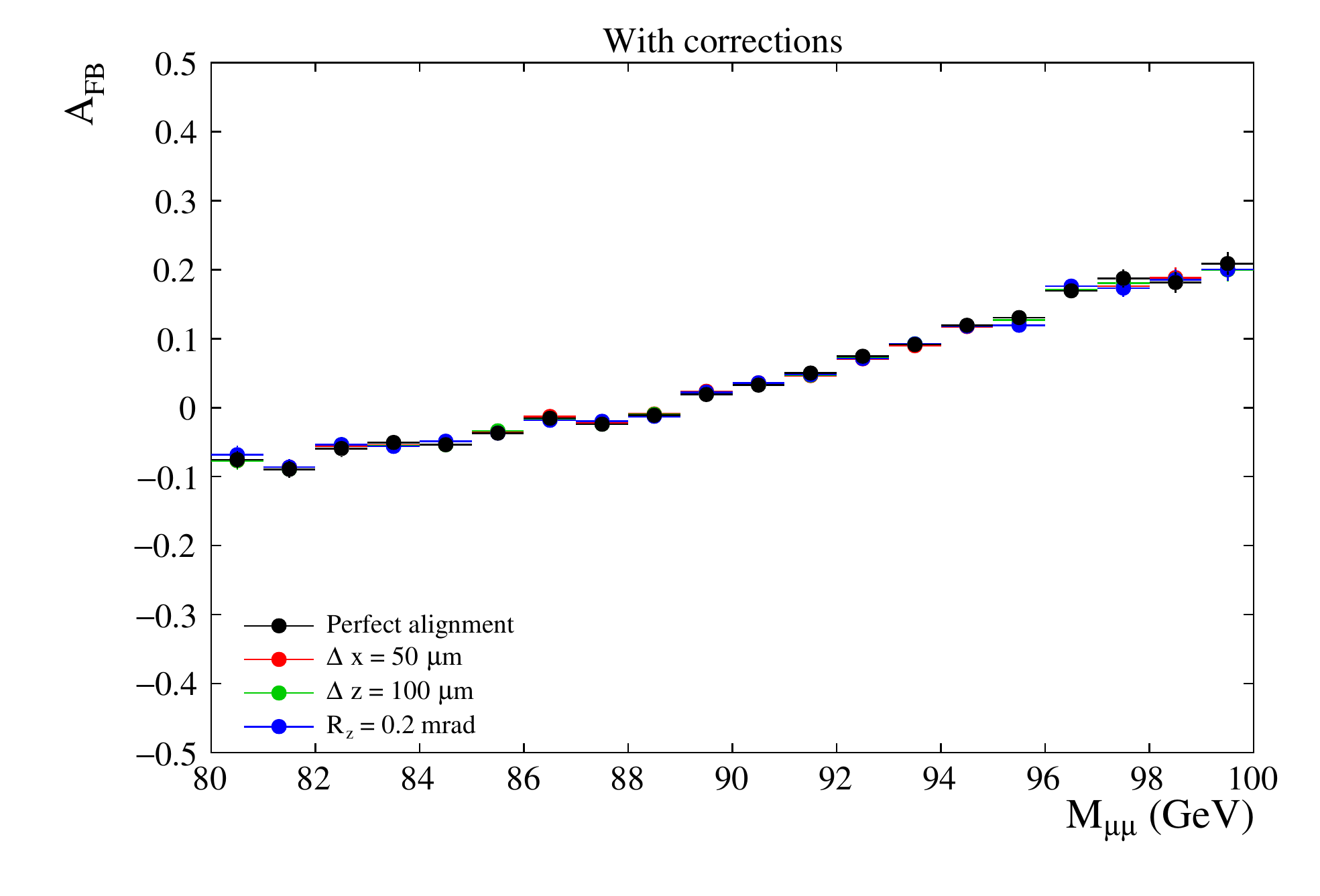}}
        \caption{The first three rows show the profiles of the mean dimuon invariant mass in bins of $\phi_+$, $\phi_-$ and $\phi_d$.
          The lower row shows the $A_{FB}$ values in bins of mass. The black points correspond to the simulation with no mis-alignment
          while the other colours correspond to the first three mis-alignment scenarios. The left and right columns correspond to the simulation before and after
          application of the pseudomass corrections, respectively.}\label{fig:summaryplots}
    \end{figure}

Maps of the curvature bias corrections in the $\eta$ and $\phi$ bins are determined for each
mis-alignment scenario, and they are presented in Fig.~\ref{fig:maps}. 
An iterative procedure is required because Eq.~\ref{eq:transf} is only accurate to leading order.
The \emph{residual}
corrections of each iterations are added to the corrections map of the previous iteration.
Iterations stop when the size of the residual corrections are zero within their statistical
errors in most $[\eta, \phi]$ bins. This happens with around two iterations.    
Figure~\ref{fig:Zmass_globTx} shows the dimuon invariant mass ($M_{\mu\mu}$)
distribution and the forward backward asymmetry in bins of $M_{\mu\mu}$ in the simulation.
The black histogram and points correspond to the simulated events before mis-alignment,
while the magenta (green) versions correspond to the mis-aligned (coherent $\Delta x = 50$\mum scenario) simulation before (after) the pseudomass corrections.
It can be seen that the mis-alignment degrades the $M_{\mu\mu}$ resolution by around 10\% and biases the $A_{FB}$
values by up to 30\% in some mass bins.
The pseudomass method successfully restores the original mass resolution and $A_{FB}$ profile.
Figure~\ref{fig:summaryplots} shows the profiles of the mean $M_{\mu\mu}$ 
as a function of the $\phi$ of the $\mu^+$ and $\mu^-$ and as a function of the angle $\phi_d$
between decay plane normal and the magnetic field direction.
In the simulated LHCb-like geometry $\phi_d$ distinguishes tracks with opposite-sign curvature in the $x-z$ plane.
The lower row of Fig.~\ref{fig:summaryplots} shows the values of $A_{FB}$ as a function of $M_{\mu\mu}$.
The left (right) columns correspond to before (after) applying the pseudomass corrections.
The black points correspond to the simulation before mis-alignment while the other three
colours correspond to the first three mis-alignment scenarios.
It can be seen that the profile of $A_{FB}$ versus $M_{\mu\mu}$ is particularly sensitive to the $\Delta x$ translation,
while the $\phi_{\pm}$ and $\phi_d$ profiles are sensitive to the $\Delta z$ translation.
The pseudomass corrections reliably resolve both of these pathologies, as can be seen in the right-hand column of Fig.~\ref{fig:summaryplots}.

\begin{figure}
    \centerline{
    \includegraphics[width=\figwidth\textwidth]{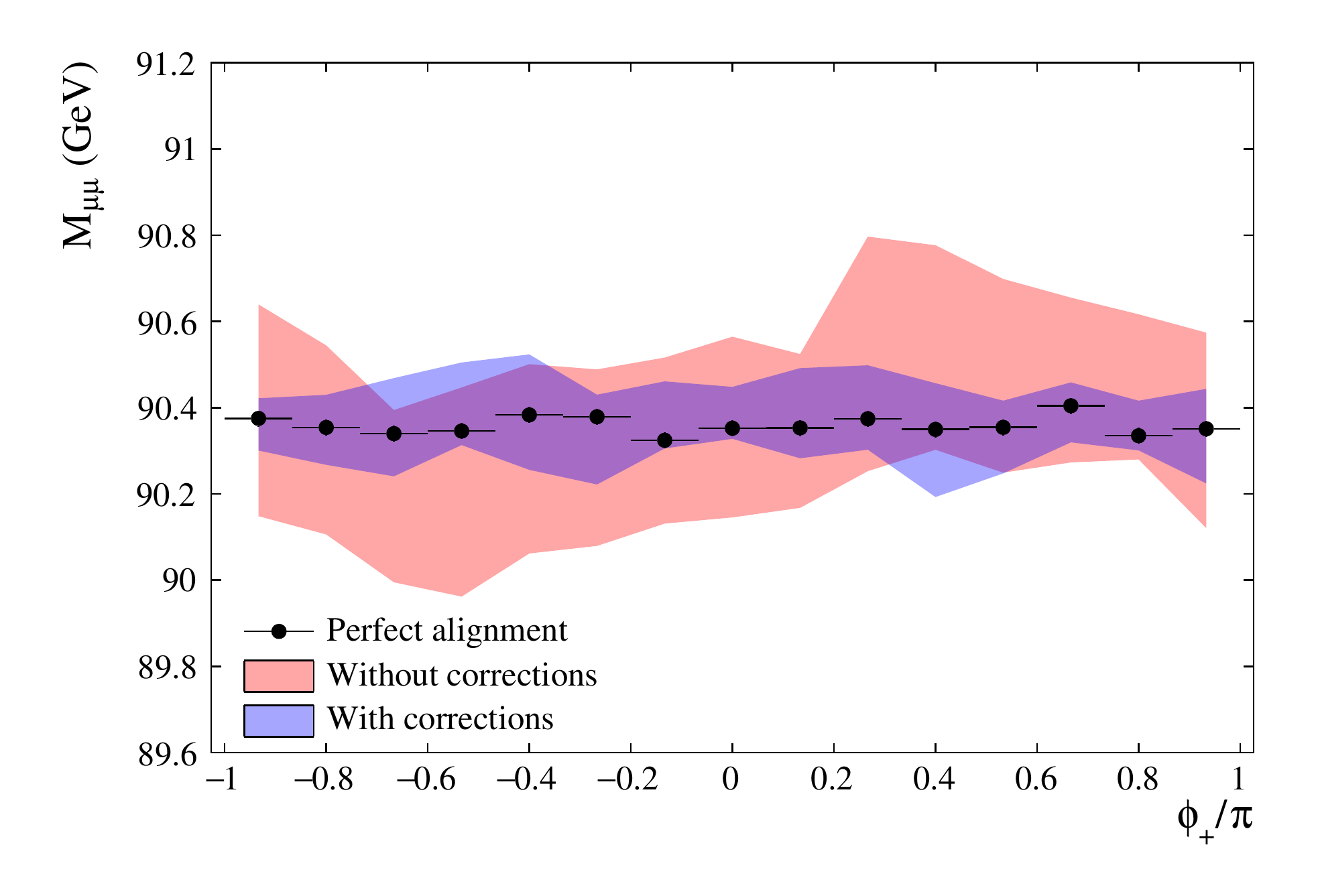}
\includegraphics[width=\figwidth\textwidth]{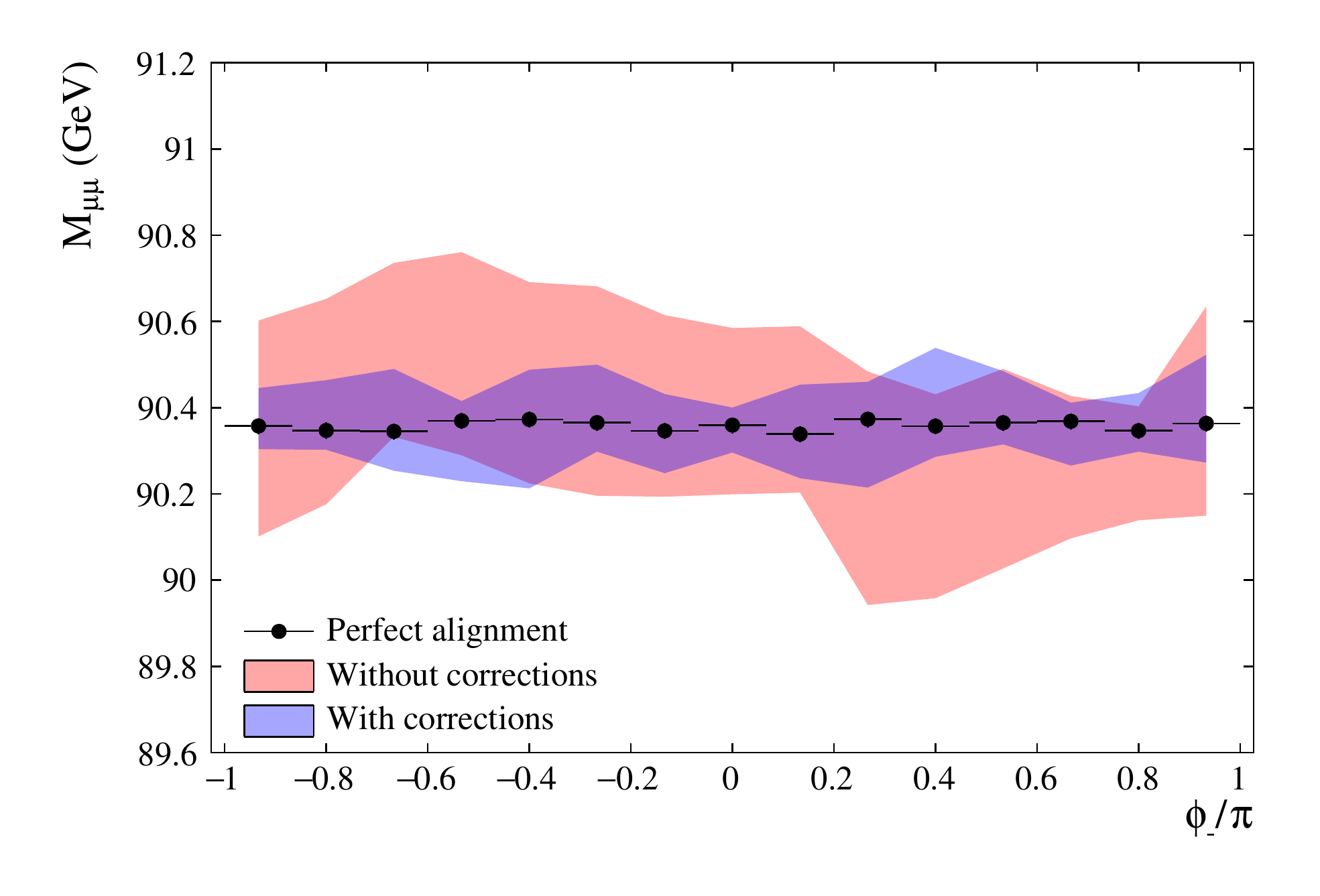}}
    \centerline{
    \includegraphics[width=\figwidth\textwidth]{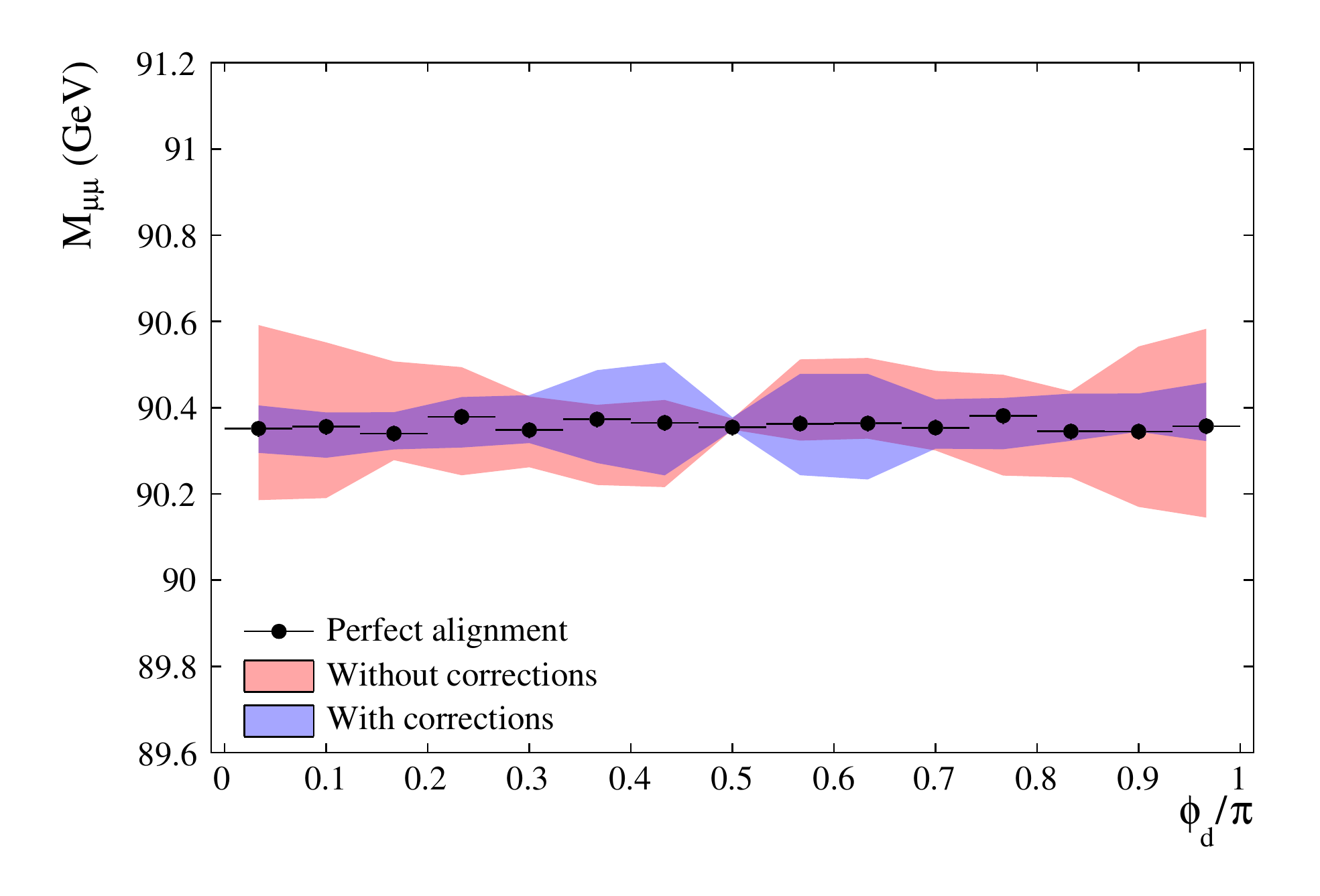}
\includegraphics[width=\figwidth\textwidth]{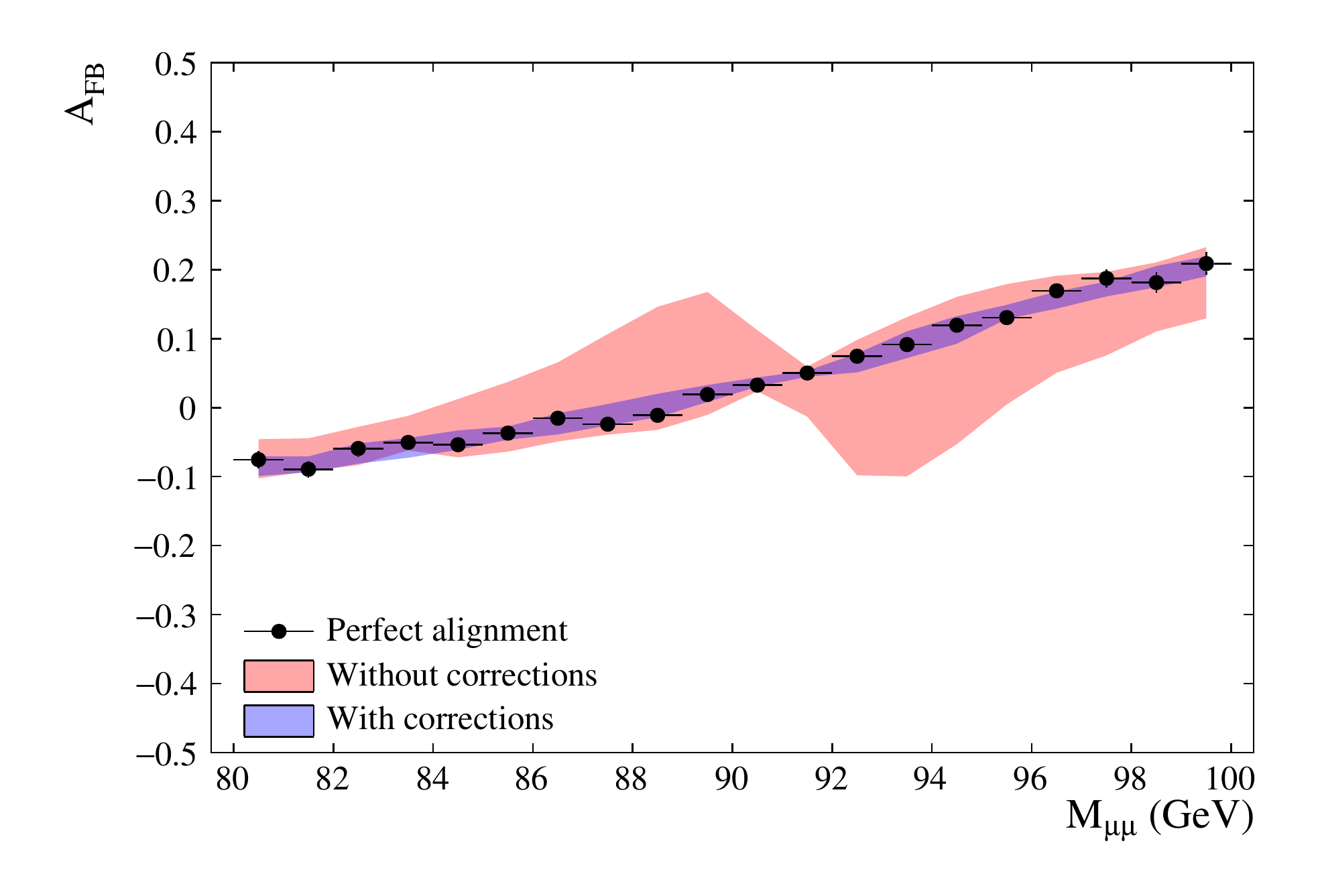}}
    \caption{Profiles of the mean dimuon invariant mass in bins of $\phi_+$, $\phi_-$ and
      $\phi_d$, and the $A_{FB}$ values in bins of mass. The black points correspond to the simulation
      before any mis-alignment. The red (blue) band represents the variations over ten toy experiments
      with the 100\mum incoherent mis-alignment scenario before (after) application of the pseudomass method.
     The centre (width) of each band corresponds to the mean (RMS) of the ten toys.        
 }\label{fig:local_toy_bands}
    \end{figure}

Figure~\ref{fig:local_toy_bands} shows the same four alignment-sensitive profiles as in Fig.~\ref{fig:summaryplots}
but considering incoherent mis-alignments along the $x$ axis.
By construction, the size of the local mis-alignment depends on
random numbers. Therefore, it is appropriate to repeat the study for multiple toy data sets, 
generated sampling the mis-alignments from different random seeds. The bands in Fig.~\ref{fig:local_toy_bands} are centered on the mean value
across 10 toys, and their width is given by their root mean square (RMS).
The red (blue) bands correspond to the values before (after) applying the pseudomass corrections to the 
mis-aligned events.
For most points the blue bands are at least a factor of two narrower than the red bands.
The reduction in the width of the band is particularly evident for the mass dependence of $A_{FB}$.
Similar results are obtained for the fifth mis-alignment scenario, corresponding
to a different binning scheme in $\eta$ and $\phi$.

It is now interesting to test the susceptibility of measurements of $m_W$ and \stw
to these mis-alignment scenarios.
For each of the first four mis-alignment scenarios, values of $m_{W^+}$, $m_{W^-}$ and
\stw are extracted from the same sample of generator level events, using a
simple single-parameter $\chi^2$ fit described in greater detail in
Section~\ref{sec:mockup}. With the assumed signal yields for LHCb Run-II~\cite{Pili}, the statistical uncertainties on $m_{W^+}$, $m_{W^-}$, and \stw are 9\,MeV, 11\,MeV and
$43 \times 10^{-5}$, respectively. 
Table~\ref{tab:EWbiases} lists, for each of the three parameters of interest, the shift in the measured value and the $\Delta\chi^2$ that is induced by the first four
mis-alignment scenarios. The overall bias for $m_W$ is obtained from the weighted average of the biases for $m_{W^+}$ and $m_{W^-}$, and is reported in a separate column in Table~\ref{tab:EWbiases}.
The largest biases in all three parameters are caused by the (coherent and incoherent) mis-alignments along the
$x$ direction. Rotations around, or translations along, the $z$ axis lead to smaller biases.
The next obvious step is to understand whether the pseudomass method is able to correct for such
biases.
\begin{table}
\begin{center}
  \caption{The biases in the values of $m_{W^+}$, $m_{W^-}$ and \stw, and the corresponding shifts in the minimum $\chi^2$ values, that are caused
    by the first four mis-alignment scenarios in an example toy measurement. 
    The overall bias in $m_W$, given by the weighted average of the biases for $m_{W^+}$ and $m_{W^-}$, is also reported.}\label{tab:EWbiases}

\resizebox{.90\textwidth}{!}{%
\begin{tabular}{ |l|c|c|c|c|c||c|c| }
    \hline
    & \multicolumn{2}{c|}{$m_{W^+}$} & \multicolumn{2}{c|}{$m_{W^-}$} &
   \multicolumn{1}{c||}{$m_{W}$} & \multicolumn{2}{c|}{\stw}\\
    \hline
    & Bias (MeV) & $\Delta\chi^2$ & Bias (MeV) & $\Delta\chi^2$ & Bias (MeV) & Bias ($\times
    10^{-5}$) & $\Delta\chi^2$\\
     \hline
     Coh. $\Delta x$ = 50\mum & -574 & 268 & +590 & 92 & -50 & +211 & 960  \\
     Coh. $\Delta z$ = 100\mum & -1.2 & 2 & -5 & 5  & -2.9 & -0.6 & 0.1  \\
     Coh. $R_z = 0.2$\,mrad & -0.4 & 0.2 & -0.1 & 1.6 & -0.3 & -4 & 3  \\
     $\Delta x = \mathrm{Gaus}(0, 100)$\mum & -101 & 52 & +98 & 79 & -11 & -65 & 28
\\
     \hline
     \end{tabular}
 }
     \end{center}
    \end{table}

%% file: Section__MockUpMeasurements.tex
\section{Impact of the pseudomass method on the measurement of electroweak observables}
\label{sec:mockup}
The sample of \Zmm events described in Section~\ref{sec:sample} is complemented by
samples of $2\times 10^{7}$ \Wmmn decays and $2\times 10^{7}$ \Wpmn decays.
Events are selected with muons in the region $2 <|\eta|< 5$, reducing both samples to around $10^7$ events.
For each of the three parameters of interest ($m_{W^+}$, $m_{W^-}$ and
\stw) 90 toy measurements are conducted.

The measurements of $m_W$ are based on template fits to the muon $p_T$ distribution of $W\to\mu\nu$ events,
while the measurements of \stw are based on template fits to $A_{FB}$ in bins of the dimuon mass in \Zmm events.
The data histograms are compared to templates where events are reweighted to emulate different $m_W$
and \stw hypotheses. 
The toy experiments are configured differently for measurements of $m_W$ and
\stw:
\begin{itemize}
\itemsep0em 
    \item In the case of $m_W$, toy data histograms are generated by randomly fluctuating the bins around the nominal
        muon $p_T$ distribution 90 times, assuming the expected LHCb Run-II yields~\cite{Pili} and Poisson
        statistics.
    \item The same procedure can not be used for the $A_{FB}$ templates, since the \Zmm events used
        for the \stw determination are also used to determine the
        pseudomass alignment corrections. It is therefore crucial to check whether this re-use of events 
        causes any bias in the determination of \stw.
        Therefore, 90 statistically independent samples are selected from the original simulated data set.
\end{itemize}

For each data histogram a single-parameter fit determines the $m_W$ or \stw
value that minimises the $\chi^2$ between the data and the templates. 
The 68\% C.L. statistical
uncertainty corresponds to a variation of $\Delta\chi^2=1$ with respect to the parabola minimum.

\begin{figure}
    \centerline{
    \includegraphics[width=\figwidth\textwidth]{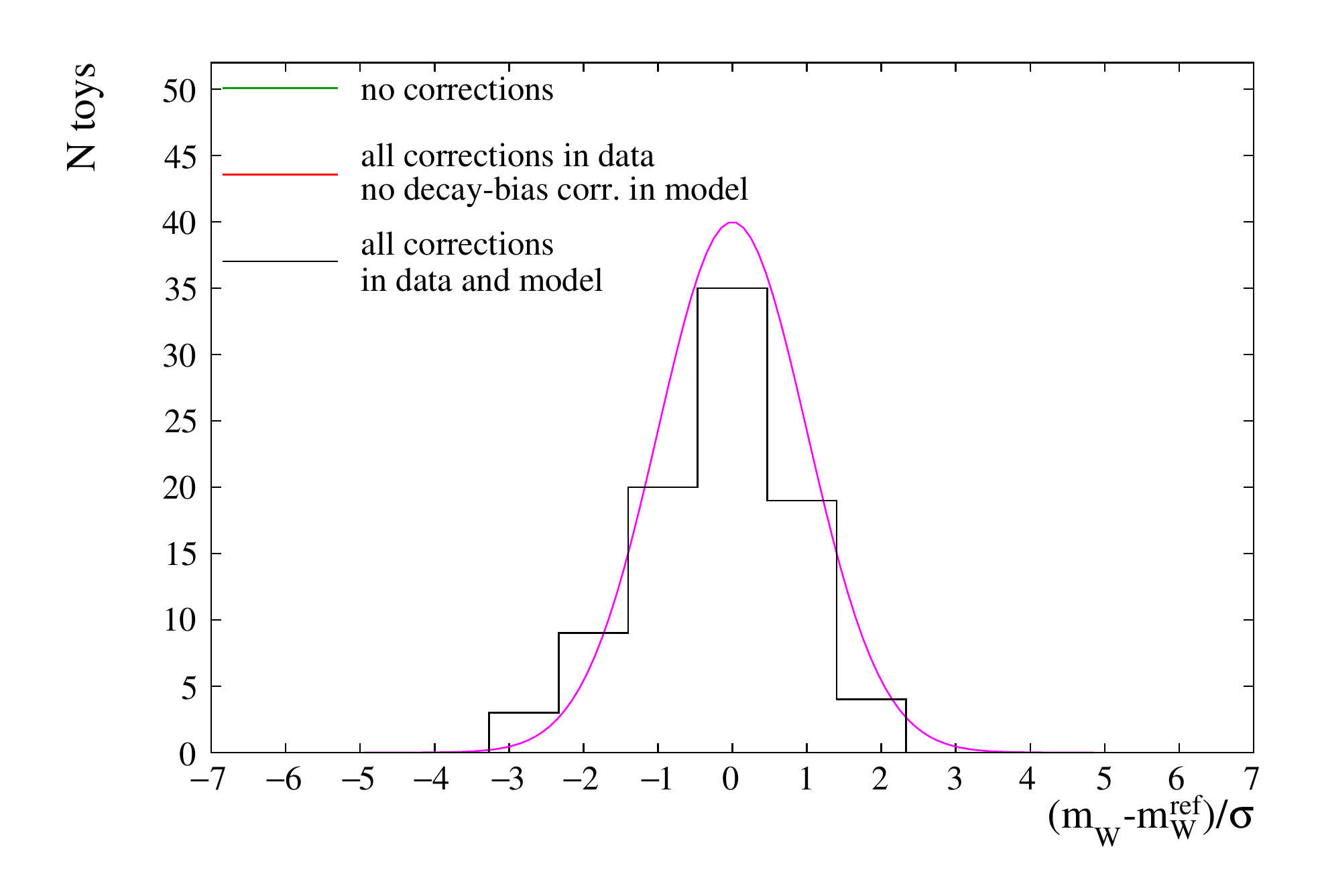}\includegraphics[width=\figwidth\textwidth]{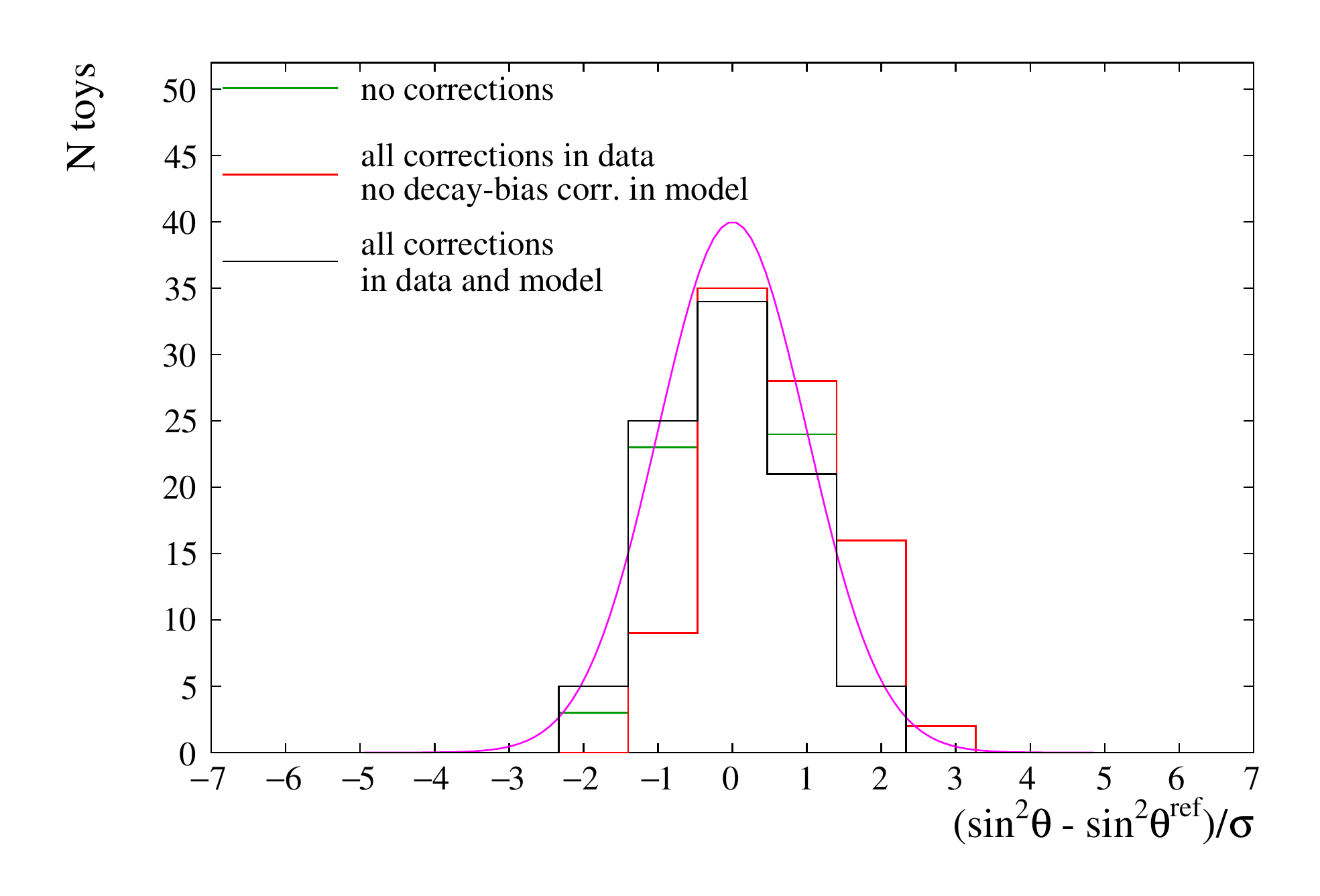}}
    \caption{The pull distributions for 
        90 toy measurements of (left) $m_{W^{+}}$  and (right) \stw. Three different scenarios are represented: 
        1) the pseudomass and decay-asymmetry bias corrections are not included in the toy
        data and template events (green); 2) the pseudomass corrections are applied to toy data and templates, but the
 decay-asymmetry bias corrections are not included in the the templates (red); 3) the pseudomass and decay-asymmetry corrections are applied to \emph{both} toy data and templates (black). Note that, in the case of $m_W$, the three distributions overlap. The expected statistical behaviour, represented by a unit Gaussian, is drawn in magenta.}\label{fig:pulls}
\end{figure}

The pull distributions for the extracted values of $m_{W^+}$\footnote{Qualitatively consistent results are seen for the $W^-$, so the figure is omitted for brevity.}, and \stw in 90 toys are shown in
Fig.~\ref{fig:pulls}. 
What is denoted as ``ref" value in the pull distributions is the nominal $m_W$ or
\stw value, which is aligned with the central template hypothesis.
A Gaussian distribution with zero mean and unit width is drawn in magenta: this is the expected distribution
over the 90 toys assuming reliable coverage of the statistical uncertainties. 
In order to study the impact of the pseudomass alignment method on the measurements of $m_W$ and
\stw, before any mis-alignment, four different scenarios are considered. 
\begin{enumerate}
\itemsep0em 
\item No pseudomass correction is included in either the toy data or template events.
    \item The pseudomass corrections are derived from the $Z$ sample and then applied to $W$ and $Z$
      events when generating the data and template histograms.
      The curvature biases are determined with Eq.~\ref{eq:transf}, i.e. without the small correction for the
      decay-asymmetry. This scenario is not shown in Fig.~\ref{fig:pulls}, but is discussed below.
    \item Same as scenario 2, but with the decay-asymmetry bias corrections (Eq.~\ref{eq:transf_corr}) applied to \emph{only} template events.
    \item Same as scenario 3, but with the decay-asymmetry bias corrections applied to \emph{both} toy data and template events.
\end{enumerate}

It can be seen in Fig.~\ref{fig:pulls} that in the first scenario both pull distributions are consistent with the ideal Gaussian functions.
A small bias is observed in the \stw pulls when the pseudomass corrections are
applied without including the decay-asymmetry bias correction in the templates (red curve), thus
introducing a discrepancy in the way the alignment corrections are applied to data and templates. 
However, the unbiased behaviour is restored after including the additional decay-asymmetry corrections.
For better visualisation, the second scenario of the list above (no decay-asymmetry bias corrections
in data and model) is not included in the plots, but 
an unbiased result is observed also in that case.

\begin{figure}
    \centerline{
    \includegraphics[width=0.45\textwidth]{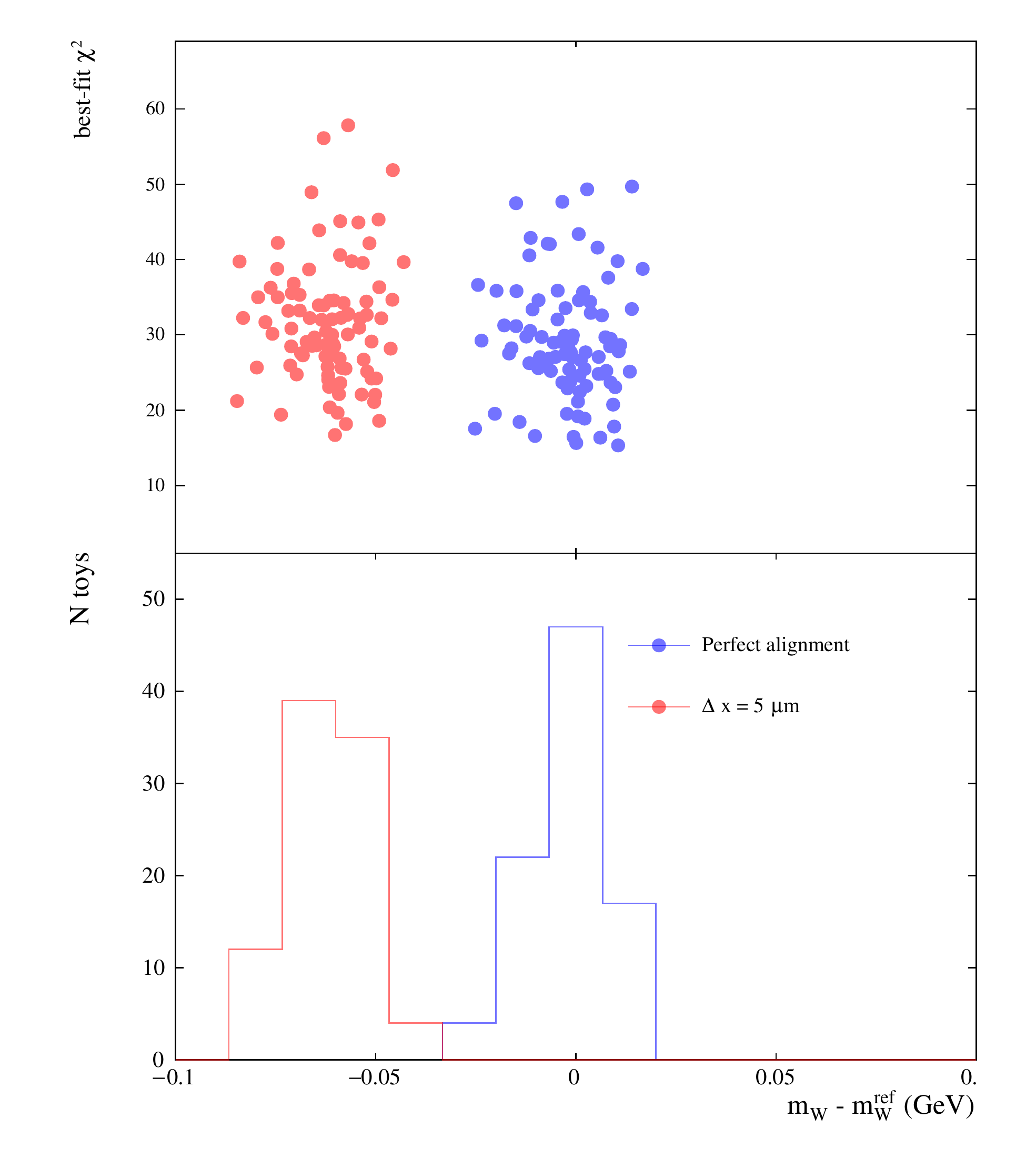}
\includegraphics[width=0.45\textwidth]{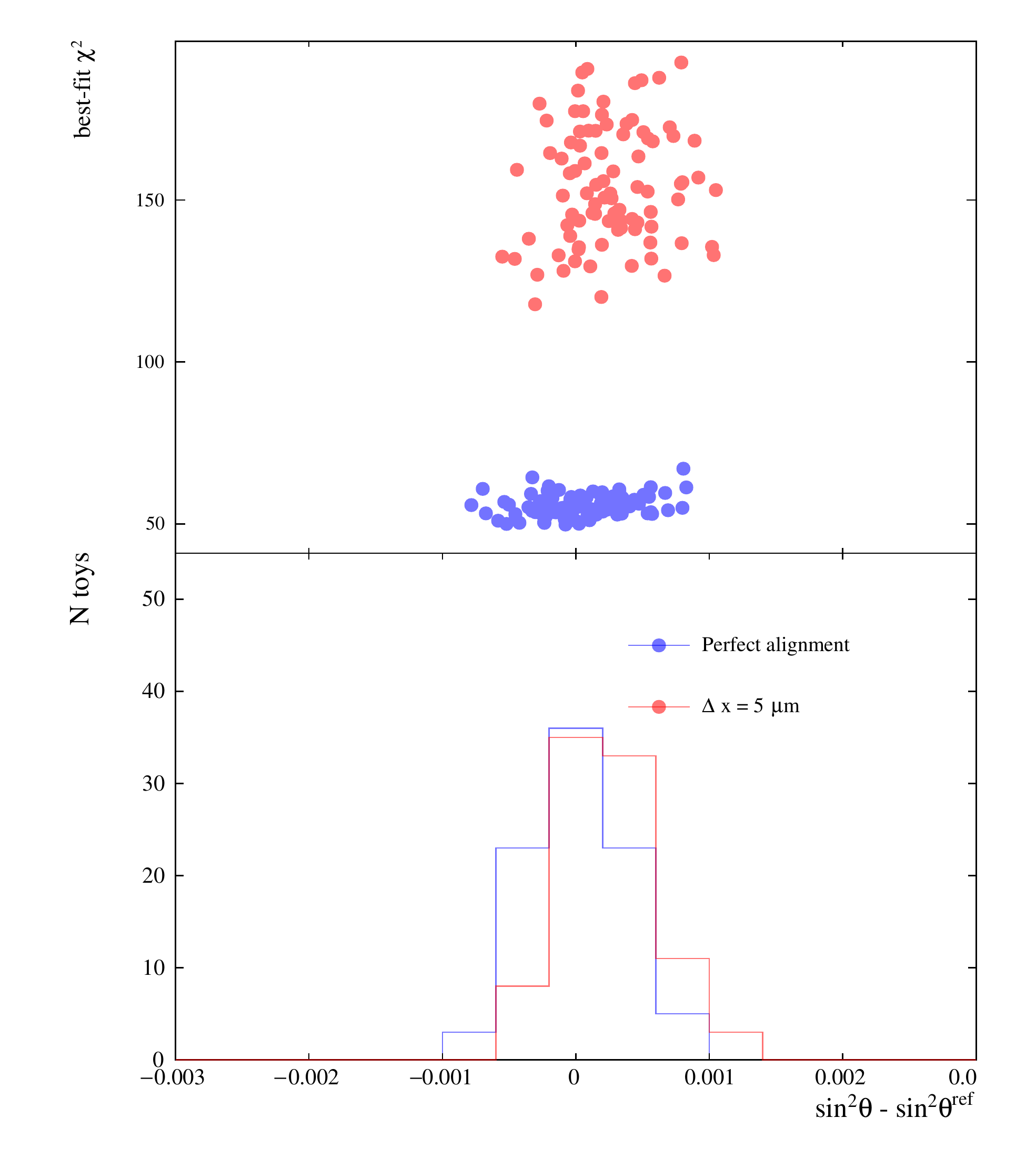}}
\centerline{
    \includegraphics[width=0.45\textwidth]{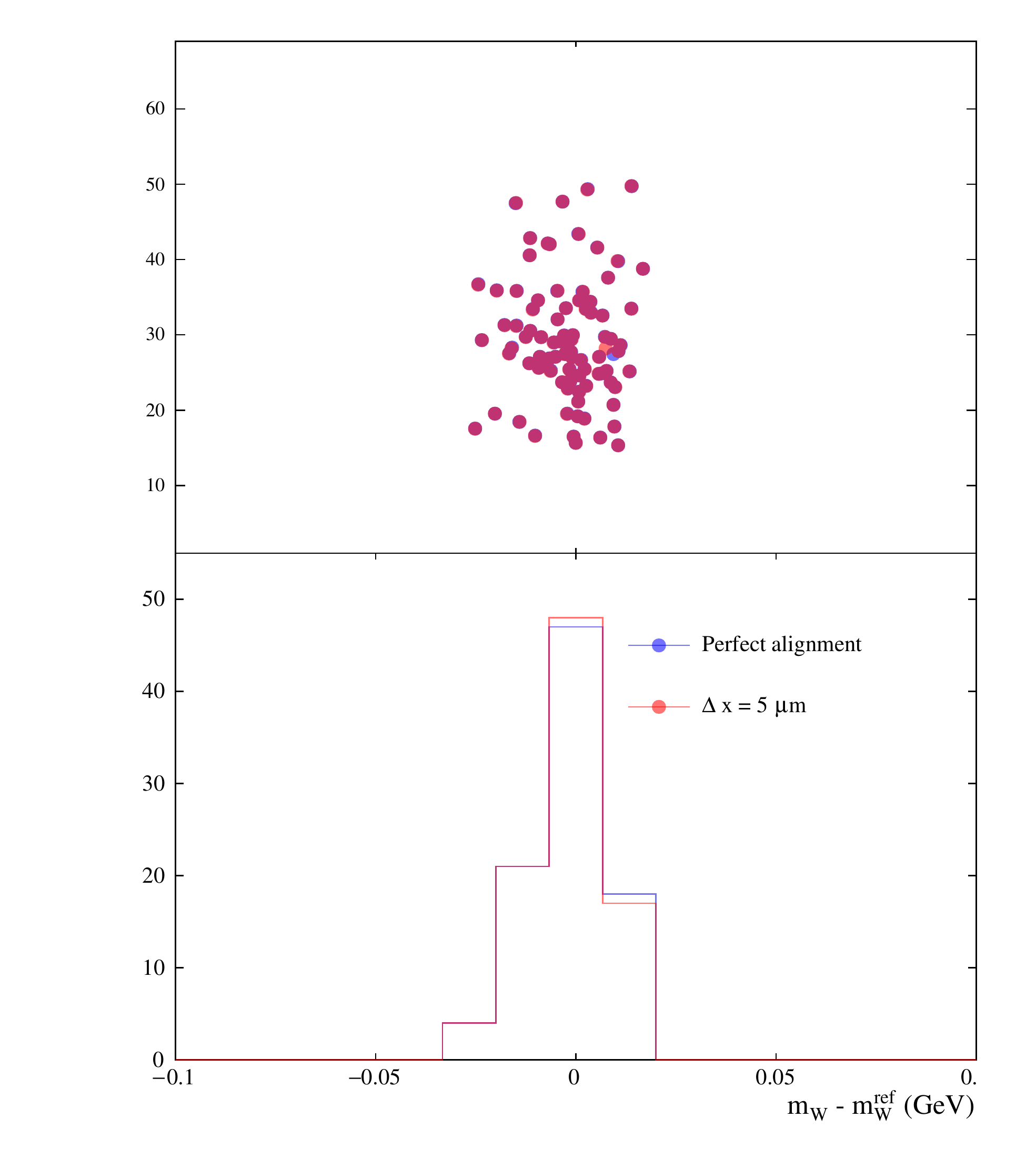}
\includegraphics[width=0.45\textwidth]{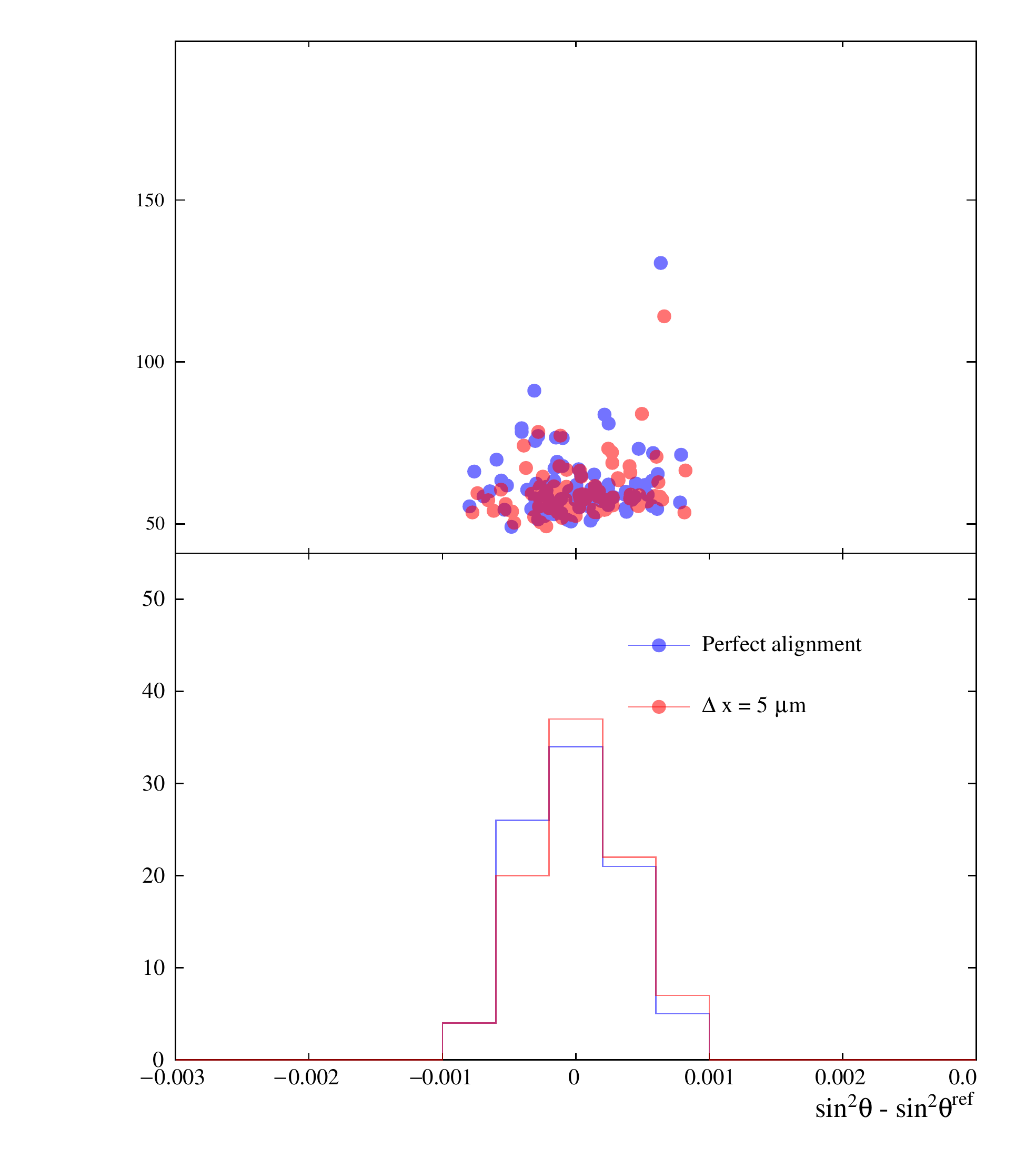}}
    \caption{The shift in the extracted (left) $m_{W^{+}}$ and (right) \stw values with respect to the reference value, 
      in 90 toys. The upper (lower) row corresponds to the simulation before (after) application of the pseudomass method.
      The distributions drawn with blue lines are obtained from unbiased data. 
        Those with red lines are obtained from toy data with a coherent 5\mum mis-alignment along $x$. 
        The first (last) row of plots show the distributions before (after) applying the pseudomass corrections. 
The upper panel of each plot shows the best-fit $\chi^2$ corresponding to the extraction of $m_W$ and \stw for the 90 toys.}\label{fig:shiftvalues}
\end{figure}

As a final test, it is interesting to check what happens to the four scenarios described above when a
detector mis-alignment is included. Figure~\ref{fig:shiftvalues} shows the shift in the extracted
$m_{W^{+}}$ and \stw values with respect to the reference value for the same 90 toys. 
The distributions drawn with blue lines are
obtained from unbiased data. Those with red lines are obtained from data with a coherent 5\mum mis-alignment along $x$. The upper (lower) row shows the results before (after) applying the pseudomass corrections. 
In order to estimate the size of the bias on $m_W$ and \stw driven by the introduced mis-alignment, the
distributions in red are fitted with a Gaussian, and the corresponding means, before and after
corrections, are reported in Table.~\ref{tab:meanpulls}. A 60\,MeV bias in the extraction of $m_{W^+}$ is estimated. Although the corresponding plot is 
not shown, we observe an opposite-sign bias of similar size for $m_{W^-}$. 
This confirms the sensitivity of the
$m_W$ measurement to small mis-alignment effects. The same size mis-alignment has a smaller effect on
the extraction of \stw, where the observed bias is within the statistical uncertainty of the measurement.  However, the corrections have an impact in
the best fit $\chi^2$ values extracted from each toy, as shown in the upper panel of each set of plots. 
\begin{table}
\begin{center}
    \caption{The mean values of the distributions shown in red in Fig.~\ref{fig:shiftvalues} (where
        a 5\mum mis-alignment along $x$ is applied to the toy data), before and
    after applying the pseudomass corrections. The biases on $m_W$ are reported in MeV.}\label{tab:meanpulls}
\resizebox{\textwidth}{!}{%
\begin{tabular}{ |c|c|c|c|c|c| }
    \hline
    \multicolumn{2}{|c|}{Bias in $m_{W^+}$ (MeV)} & \multicolumn{2}{c|}{Bias in $m_{W^-}$ (MeV)} &
    \multicolumn{2}{c|}{Bias in \stw ($\times 10^{-5}$) }\\
    \hline
    Before\,(After) corr. & $\sigma_{\text{stat}}$ on $m_{W^+}$ & Before\,(After) corr. & $\sigma_{\text{stat}}$ on $m_{W^-}$ & Before\,(After) corr. & $\sigma_{\text{stat}}$ on \stw \\
     \hline
     -62\,(-1)  & 9 & +56\,(0.7) & 11 & +26\,(9) & 43  \\
     \hline
     \end{tabular}
 }
     \end{center}
    \end{table}

%% file: Section__Conclusion.tex
\section{Conclusion}
The measurements of, for example, $m_W$ and \stw, using muonic decays of weak bosons at hadron colliders
are susceptible to biases in the measurement of muon momenta.
A particular concern is curvature biases, caused by mis-alignments of the tracking detector elements, that depend on the sign of the particle charge.
It is proposed to use the ``pseudomass method", which is introduced in this paper, to
determine corrections for charge-dependent curvature biases using \Zmm decays.
The method is validated using simulated $pp \to \Zmm$ events with the LHC Run-II centre-of-mass energy
of 13~TeV. A simplified model of a detector with a similar geometry to the LHCb experiment is used.
This approach has the advantage of being less dependent on
assumptions about the kinematics of the $Z$ boson decays than other methods present in the literature.
A small correction for an effect of the forward-backward asymmetry in \Zmm decays is required
but, importantly, the curvature biases can be determined with limited sensitivity to assumptions about the value of \stw.
The method is tested against several simplified mis-alignment configurations.
With pseudo-experiments using simulated \Zmm and $W\to\mu\nu$ decays it is demonstrated that the proposed method
can be reliably applied in measurements of $m_W$ and \stw in the presence of these simplified mis-alignments.

%% file: Section__Acknowledgements.tex
\section*{Acknowledgements}
We thank T.~Wyatt for making the authors aware of the introduction, and usage of, the pseudomass variable in an analysis of data from the D0 experiment,
which prompted the present study.
We thank W.~Hulsbergen for helpful suggestions that improve the clarity of our description of the pseudomass method and P.~Ilten for his support with Monte Carlo generators.
We also thank N.~Tuning and L.~Sestini for the useful comments, as well as O.~Lupton, S.~Farry, R.~Hunter, M.~Ramos~Pernas, A.~Chadwick, H.~Yin and M.~Xu for interesting discussions in the context of analyses of $W$ and $Z$ boson decays at LHCb. WB is supported by an Imperial College Research Fellowship. MV is supported by the grants ERC-CoG-865469 SPEAR and STFC-ERF-ST/N004892/2. 